\documentclass[fleqn,usenatbib]{mnras}

\usepackage{amsmath}
\usepackage{url}
\usepackage{amsfonts}
\usepackage{amsbsy}
\usepackage{verbatim}
\usepackage{array}
\usepackage{amssymb}
\usepackage{amsbsy}
\usepackage{graphicx,xcolor,color,subfig}
\usepackage{soul}

\renewcommand{\u}{\ensuremath{\mathbf{u}}}
\renewcommand{\d}{\ensuremath{\partial}}
\newcommand{\w}{\ensuremath{\mathbf{w}}}

\newcommand{\ex}{\ensuremath{\mathbf{e}_{x}}}

\newcommand{\ez}{\ensuremath{\mathbf{e}_{z}}}

\title[3D simulations of the COS]{Local three-dimensional simulations of the convective overstability in protoplanetary discs}
\author[R.J.~Teed, H.N.~Latter]{Robert J.~Teed\thanks{E-mail:
   Robert.Teed@glasgow.ac.uk}$^{1,}$, Henrik N.~Latter$^{2}$ \\
$^{1}$ School of Mathematics and Statistics, University of Glasgow, University Place, Glasgow G12 8SQ, UK\\
$^{2}$ DAMTP, University of Cambridge, CMS, Wilberforce Road,
Cambridge CB3 0WA, UK}

\begin{document}

\maketitle

\begin{abstract}
At certain radii protoplanetary discs may sustain a form of oscillatory convection 
(`convective
overstability'; COS) due to localised adverse entropy
gradients. The resulting hydrodynamical activity can produce coherent structures, such as zonal flows and vortices, that may concentrate solid material and aid their further coagulation. In this paper we extend previous axisymmetric runs by performing local three-dimensional simulations of the COS, using the code SNOOPY. 
As parameters are varied, we characterise how the various axisymmetric COS saturated states are transformed in 3D, while also tracking their interrelationship with the subcritical baroclinic instability. In particular, at low Reynolds number (Re) our 3D simulations exhibit similar weakly nonlinear and wave turbulent states to our earlier axisymmetic runs. At higher Re, but low Peclet number (Pe), we obtain bursty cycles involving the creation of zonal flows, the subsequent development of planar vortices, and their destruction by elliptical instability. For larger Pe, however, zonal flows can persist, alongside weaker more elongated vortices. These results further reveal the diversity of the COS's behaviour, and show that solid accumulation via COS-induced vortices may not be straightforward. 
\end{abstract}


\section{Introduction}

For the most part, protoplanetary (PP) discs are insufficiently ionised to support magnetorotational turbulence, and accretion is thought to be driven by laminar magnetic winds, at least at radii greater than $\sim 1$ AU \citep{turner14,lesur21}. However, research over the last 15 years has revealed that these `dead' regions are subject to a fascinating array of hydrodynamical instabilities that typically generate structures (zonal flows and vortices) as part of their saturation \citep{fl19, lu19,lesur23}. They, in turn, may impact on solid dynamics and ultimately planet formation \citep{draz23}.
Of these instabilities, we focus on the convective overstability (COS), a form of oscillatory convection with a double-diffusive character \citep{KH14,lyr14, lat16}, which arises in pockets of radially decreasing entropy \citep[e.g., ice lines, planetary gaps, and the dead/active zone boundary; see discussion in][hereafter TL21]{tl21} 

Simulations in Boussinesq shearing boxes show that the axisymmetric COS saturates in remarkably diverse states, depending on the strength of the entropy gradient and viscosity (TL21). In particular, as the Reynolds number (Re) is increased, the saturation proceeds through (a) a weakly nonlinear state, (b) inertial wave turbulence, (c) intermittent zonal flows, and (d) persistent zonal flows. 
Zonal flows are usually accompanied by elevator flows, which may be physical but are artificially enhanced in shearing boxes.
Additionally, 3D simulations reveal that the COS produces planar vortices \citep{lyr14}, which can collect dust \citep{raet21,lyra24}. Recently, researchers have begun to explore global models of the COS, which extend, and complicate, some of the results discovered in local boxes \citep{Lehm24,Lehm25}.

Related but distinct to the COS is the subcritical baroclinic instability (SBI), a nonlinear process confined to the disc plane \citep{klahr03,peter07,lp10}. The instability mechanism relies on a finite-amplitude vortical seed, and thus amplifies and reshapes pre-existing vortices. The magnitudes of thermal diffusion and the entropy gradient select which vortices are preferentially excited (e.g., efficient diffusion leads to larger vortices); meanwhile, the SBI reduces their aspect ratio \citep[][hereafter LP10]{lp10}. The relationship between the COS's and SBI's nonlinear saturation is yet to be explained.

In this paper we extend our survey of axisymmetric COS states (TL21) to three dimensions, using Boussinesq shearing boxes and the code SNOOPY. Key questions we seek to answer include (a) do the weakly nonlinear and wave turbulent states remain effectively axisymmetric, (b) do zonal flows become unstable to shear instability and thus the creation of vortices, (c) what are the properties and lifetimes of vortices so created, and (d) how does the SBI interact with these vortices? Given the numerical cost of 3D simulations, we are unable to range across as wide a parameter space as in TL21. Instead, we focus mostly on single values of the Richardson and Peclet numbers and vary Re.  

We find that that three-dimensional effects impact only on the states that exhibit zonal flows. At lower Peclet numbers, these periodically break down via non-axisymmetric shear instability; the vortices that emerge from this process also break down, but via elliptical instability. The process repeats on a timescale of roughly 30-40 orbits and the SBI appears to both enhance and enlarge the vortices so formed. At higher Peclet numbers this is not the case: persistent zonal flows are possible, as in axisymmetry, albeit co-existing with weaker more elongated vortices. We expect the highly disordered dynamics revealed by our simulations to complicate the commonly held picture that the COS simply forms vortices and that these reliably accumulate solids.

\section{Methods}

\subsection{Physical model and governing equations}

As in TL21, we
employ the Boussinesq shearing
box \citep{glb65, hawley95, lp17}.
Its governing equations are
\begin{align} \label{GE1}
&\d_t \u + \u\cdot\nabla\u = -\frac{1}{\rho}\nabla P -2\Omega \ez\times
\u \notag\\ 
& \hskip1.5cm + 2q\Omega\, x\,\ex -N^2\theta\,\ex  +\nu\nabla^2\u, \\
& \d_t\theta + \u\cdot\nabla\theta = u_x + \xi\,\nabla^2\theta,\label{GE2} \\
& \nabla\cdot \u = 0, \label{GE3}
\end{align}
where $\u$ is the fluid velocity, $P$ is pressure, $\rho$ is the
(constant) background density, and $\theta$ is the buoyancy variable. The fixed orbital frequency of the box is $\Omega$, while the
shear parameter of the sheet is denoted by $q$. The buoyancy frequency issuing from the radial
stratification is denoted by $N$. 
The thermal diffusivity is $\xi$, and the kinematic viscosity is $\nu$.

In addition to $q$, the system can be specified by three other
dimensionless parameters.
The `$R$' number measures the relative strength of the (unstable)
radial stratification to the stabilising angular momentum gradient:
\begin{equation}
R = - \frac{N^2}{\kappa^2},
\end{equation}
where $\kappa^2= 2(2-q)\Omega^2 $ is the squared epicyclic frequency.
The relative importance of the diffusivities is measured by the
Peclet and Reynolds numbers
\begin{equation} \label{Renum}
\text{Pe}= \frac{L^2\kappa}{\xi}, \qquad \text{Re}=\frac{L^2\kappa}{\nu},
\end{equation}
where $L$ is a characteristic outer lengthscale, which we take to be the height of our box. For a fuller discussion of the model, its assumptions, and its parameters see \citet{lat16} and TL21.

\subsection{Numerical techniques}

\subsubsection{Code and set-up}

Numerical simulations are performed with the code, SNOOPY \citep{les05,les07}, which solves the shearing box equations
using a pseudo-spectral method based on a shearing wave decomposition. Further details on the code are also given in TL21.

Our 3D domain has dimensions $L_x\times L_y\times L_z$ and is periodic in all three coordinates. 
The majority of our simulations are performed with $L_x=2L$, $L_y=4L$, $L_z=L$ and a grid resolution of $512\times1024\times256$ points.

To complement our 3D runs we perform 
2D simulations similar (though not identical; see below) to LP10 where $L_x=2L$ and $L_y=4L$ with grid resolution $512\times1024$. These are labelled `2Dxy'.

Our units are set such that $L=1$, $\Omega=1$, and $\rho=1$, matching TL21. This leads to a difference in our 2Dxy set-up to that of LP10 where units were chosen with $\Omega=2/3$ (or, equivalently, $q=1$). 

\subsubsection{Parameter values and initial conditions}

In all runs, the disc is
Keplerian and so $q=3/2$, which leaves three dimensionless parameters:
 $R$, Pe, and Re. Owing to the vastly increased computational requirements to perform 3D simulations over the axisymmetric set-up of TL21, we do not sweep through a large portion of the parameter space, but rather focus on representative behaviours as a single parameter varies, namely Re. To `speed up' the dynamics, we select $R=0.1$ and mostly take Pe$=4\pi^2\simeq40$. The Reynolds number varies from about $10^3$ to $10^5$. For a fuller discussion on the implications of these parameter choices, see TL21.

 Our 3D runs are seeded with small amplitude noise of a size roughly $|\mathbf{u}|\sim10^{-5}$. In contrast, our SBI runs in 2D require finite-amplitude initial conditions. Following LP10, we initialise large scale modes (above the cut-off lengthscale of $\ell_\text{cut}=1/3$) with 2D noise of amplitude $|\mathbf{u}|\sim0.6$.

\subsubsection{Diagnostics}

As well as the velocity field, $\u$, it is informative to compute the vorticity, $\w=\nabla\times\u$,
and, in particular, its vertical component, $w_z$, which can be obtained by calculating derivatives in spectral space.
Beside the fields themselves, our main diagnostics are the box-averaged kinetic energy and its directional components:
\begin{align}
E_K &= \langle \rho|\u|^2 \rangle,\qquad \\
E_{K_i} &= \langle \rho u_i^2 \rangle\qquad (\text{ for } i\in\{x,y,z\})
\end{align}
Angle brackets denote an average over the full spatial domain 
and subscripts signify an average over one or two of the three coordinates. 
For example, $\langle\cdot\rangle$ and $\langle\cdot\rangle_{yz}$ represent 
a full box average and an average over $y$ and $z$ only, respectively.

\section{Results}

We opt to mostly focus on the sequence of states that emerge as Re varies, with fixed values of $R$ and Pe (TL21). The choice $R=0.1$ and Pe$=4\pi^2\simeq40$ supplies a particularly clear sequence (see Fig.~16 of TL21): weakly non-linear waves, wave turbulence, wave turbulence and zonal flows, and persistent zonal flows.
This allows us to illustrate the merits of the axisymmetric set-up but also the key differences in 3D; further computational resources would be required for a more comprehensive study varying $R$ (and Pe more widely). The results section is structured around the different dynamical states we achieve, with most attention devoted on the novel cycles of zonal flows and vortices.

\subsection{Weakly nonlinear (WNL) state}

\begin{figure}
    \centering
    \includegraphics[width=0.9\linewidth]{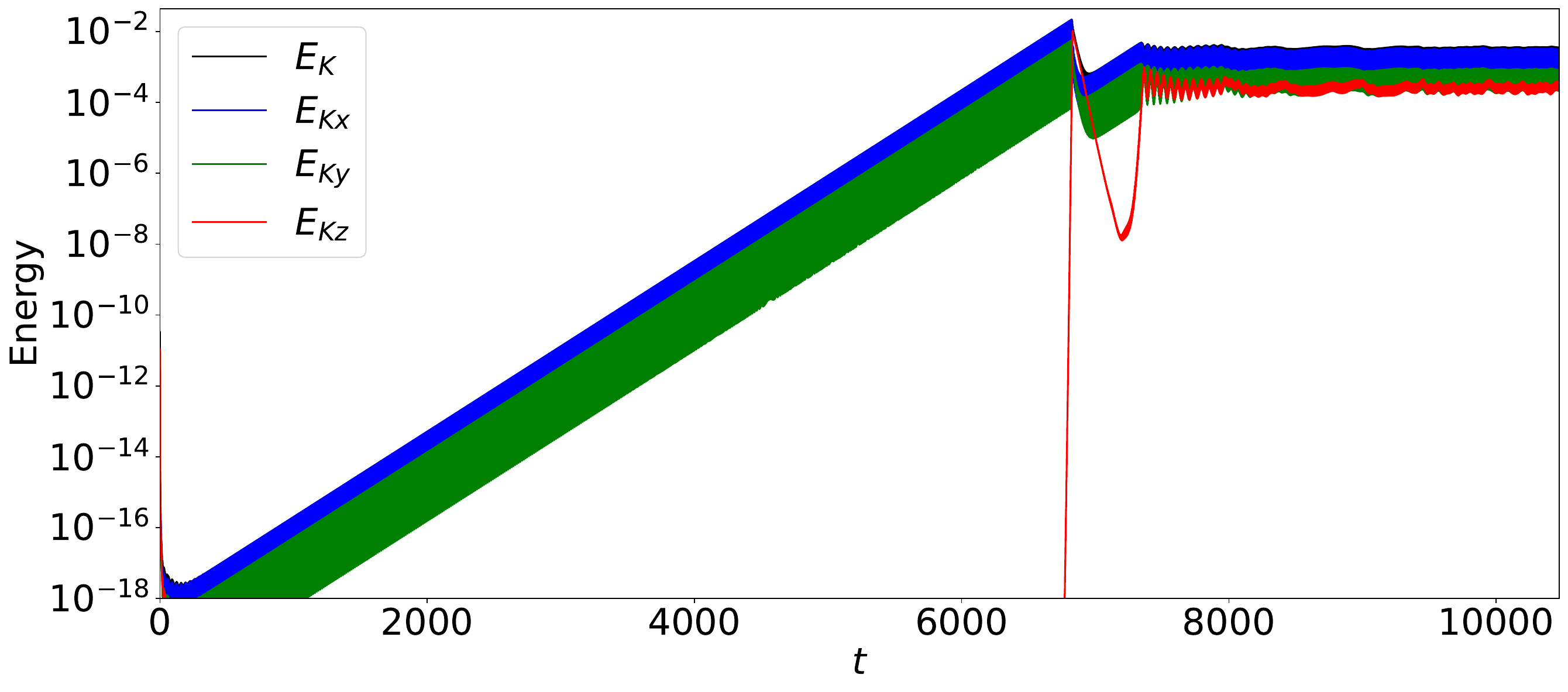}
    \caption{Time series of the energy, and its directional parts, for a solution in the weakly nonlinear regime $(\text{Re}=10^{3.25}, \text{Pe}=4\pi^2)$.}
    \label{fig:WNL-energy}
\end{figure}

Starting from Re$\simeq10^{3.25}$, as in TL21, axisymmetric instability occurs, corresponding to an exponential growth of the kinetic energy dominated by the radial flow and, to a lesser extent, the azimuthal flow. (The vertical component remains insignificant until saturation.) The dominant mode in this phase is the $k_x=k_y=0$, $k_z=2\pi$ epicyclic oscillation, as in TL21. Fig.~\ref{fig:WNL-energy} shows this behaviour, which is similar across all simulations. For this reason, we do not present full time series for the remainder of this work. Note that the curves in Fig.~\ref{fig:WNL-energy} include oscillations near the orbital frequency that generate large fluctuations in energy in narrow time windows. When displayed over a long time series, the result is a thick block of colour!

\begin{figure}
    \centering
     (a)\hskip -2mm \hspace{2mm}\vspace{3mm} \includegraphics[width=0.9\linewidth]{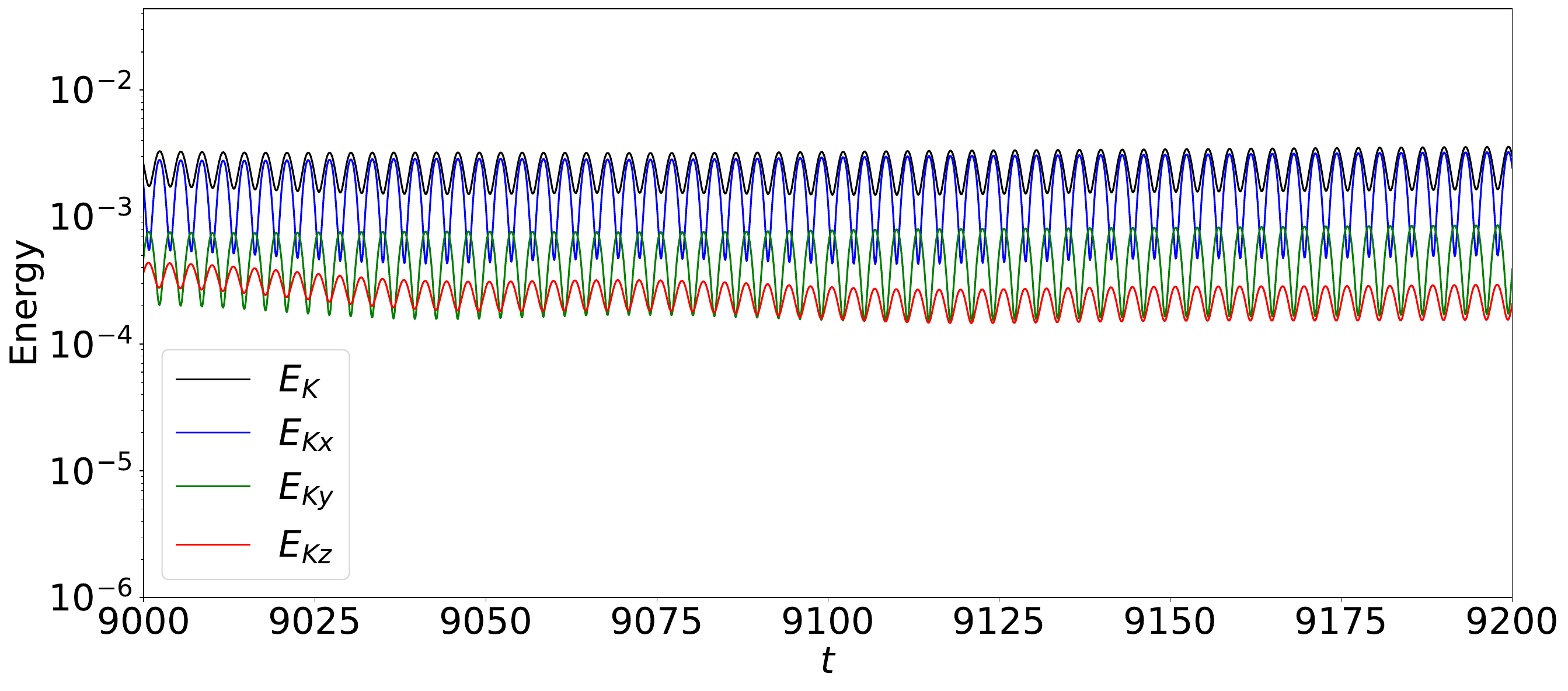}
    (b)\hskip -2mm \hspace{2mm} \vspace{3mm}{\label{fig:WNL-ux-xz}\includegraphics[width=0.9\linewidth]{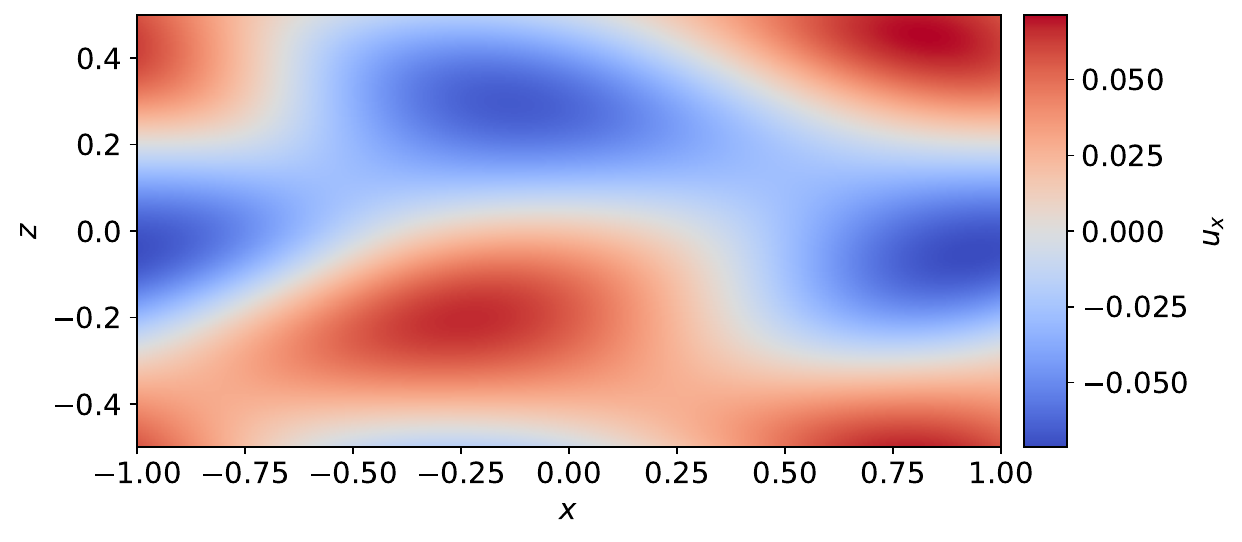}}
    (c)\hskip -2mm \hspace{2mm} \vspace{3mm}{\label{fig:WNL-uz-xz}\includegraphics[width=0.9\linewidth]{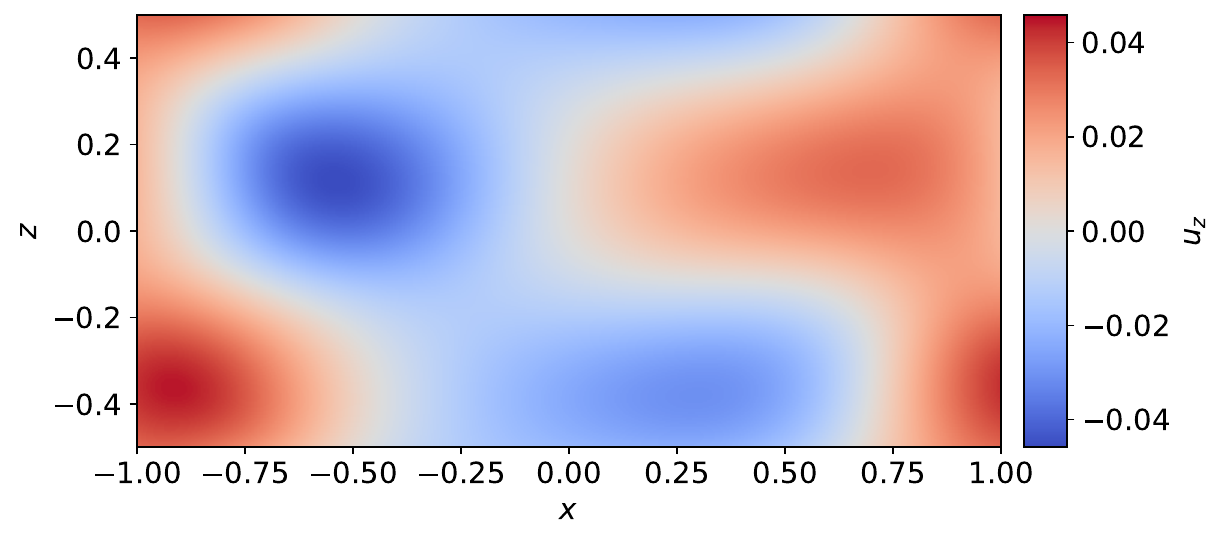}}

    (d)\hskip -2mm \hspace{2mm} {\label{fig:WNL-uz-xy}\includegraphics[width=0.42\linewidth]{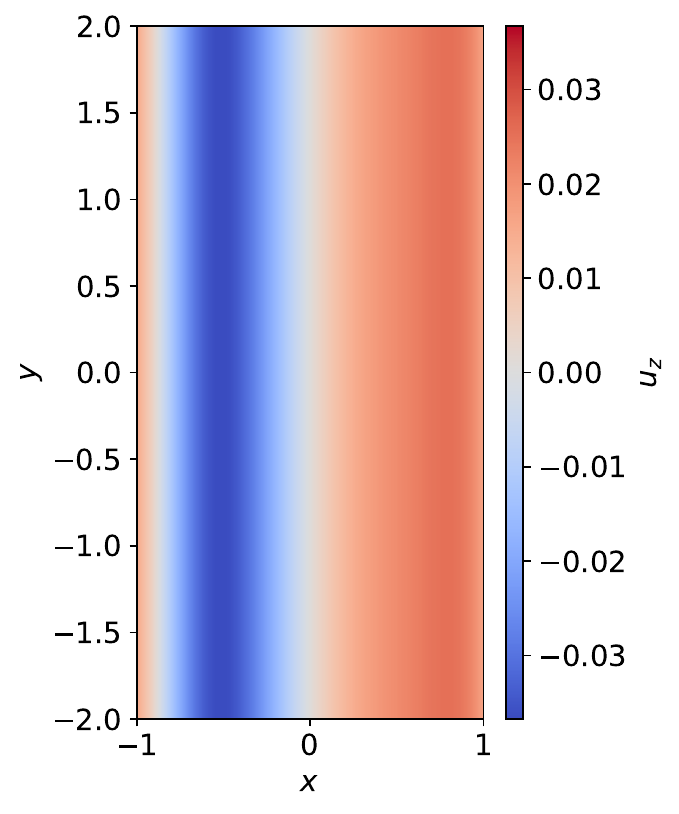}}
    (e)\hskip -2mm \hspace{2mm} {\label{fig:WNL-wz-xy}\includegraphics[width=0.42\linewidth]{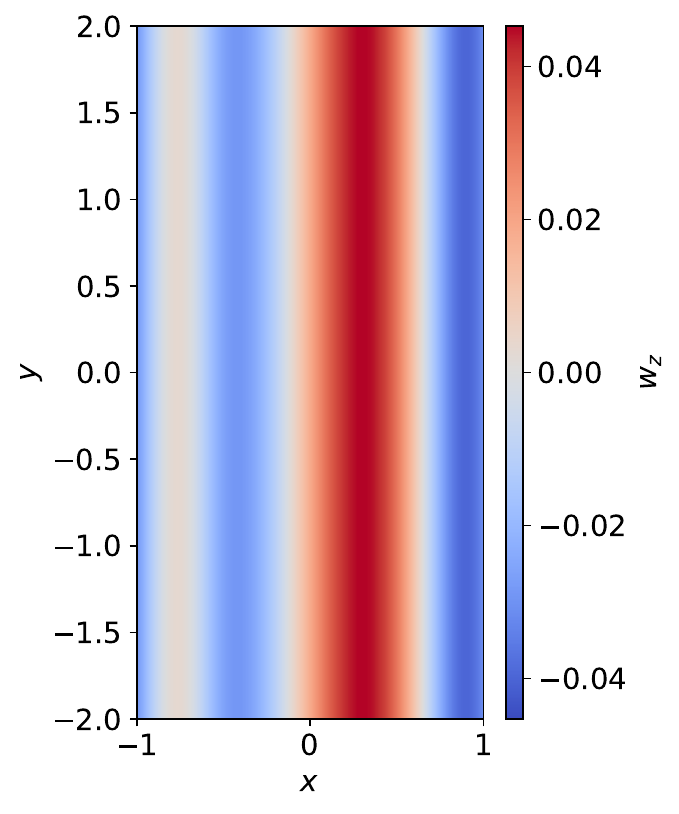}}
    \caption{Plots for a solution in the weakly nonlinear regime $(\text{Re}=10^{3.25}, \text{Pe}=4\pi^2)$. (a) Partial time series of the energy, and its directional parts. (b-e) Plots of components of the velocity field at $t=1.04\times10^4$: (b) $u_x$ in an $xz$-slice (at $y=0$);
    (c) $u_z$ in an $xz$-slice (at $y=0$); (d) $u_z$ in an $xy$-slice (at $z=0$); (e) $w_z$ in an $xy$-slice (at $z=0$).}
    \label{fig:WNL}
\end{figure}

The growth rate of the dominant linear COS mode at Re$=10^{3.25}$ is weak, leading to a lengthy period of linear growth before a weakly nonlinear saturated state emerges at $t\gtrsim 7500$, oscillating near the orbital frequency (Fig.~\ref{fig:WNL}(a)). The majority of the kinetic energy continues to be associated with radial motion, and the primacy of the fastest growing axisymmetric COS mode remains, as a spatial Fourier decomposition reveals (not shown). However, Fig.~\ref{fig:WNL}(a-c) suggest a small number of additional axisymmetric modes contribute to the saturated state. Modes with radial wavelength $2$ and $1$, and vertical wavelength $1$ are the most pronounced, which we identify as slower growing COS modes. A number of shorter-scale modes, presumably in resonance with the dominant COS modes, achieve smaller amplitudes but are likely key in redistributing energy input by the instability and bringing the system to saturation. This weakly nonlinear state, composed of a small number of interacting wave modes, is qualitatively similar to those explored in TL21, though more complicated than the example they exhibited (which consisted of mainly three modes). 

Finally, we demonstrate the strong axisymmetry of this state by plotting snapshots of the vertical velocity (Fig.~\ref{fig:WNL}(d)) and vertical vorticity (Fig.~\ref{fig:WNL}(e)) in the $xy$ plane. Note that these structures are unsteady, and in fact oscillate and propagate on the fast orbital timescale. This, given their relatively weak velocities, means they are not easily subject to non-axisymmetric shear instability (see discussion later). In summary, for these low Re, 3D simulations and axisymmetric simulations of the COS are practically the same.

\subsection{Wave turbulent state}

An increase of the Reynolds number leads to saturated states dominated by inertial wave turbulence. Within our simulation suite we found the wave turbulent regime occurs between Re$=10^{3.5}$ and Re$=10^4$ (with Pe$=4\pi^2$). A larger growth rate compared to the WNL state leads to saturation for $t\gtrsim 1700$ when Re$=10^{3.5}$.

\begin{figure}
    \centering
    (a)\hskip -2mm \hspace{2mm} \vspace{3mm}\includegraphics[width=0.9\linewidth]{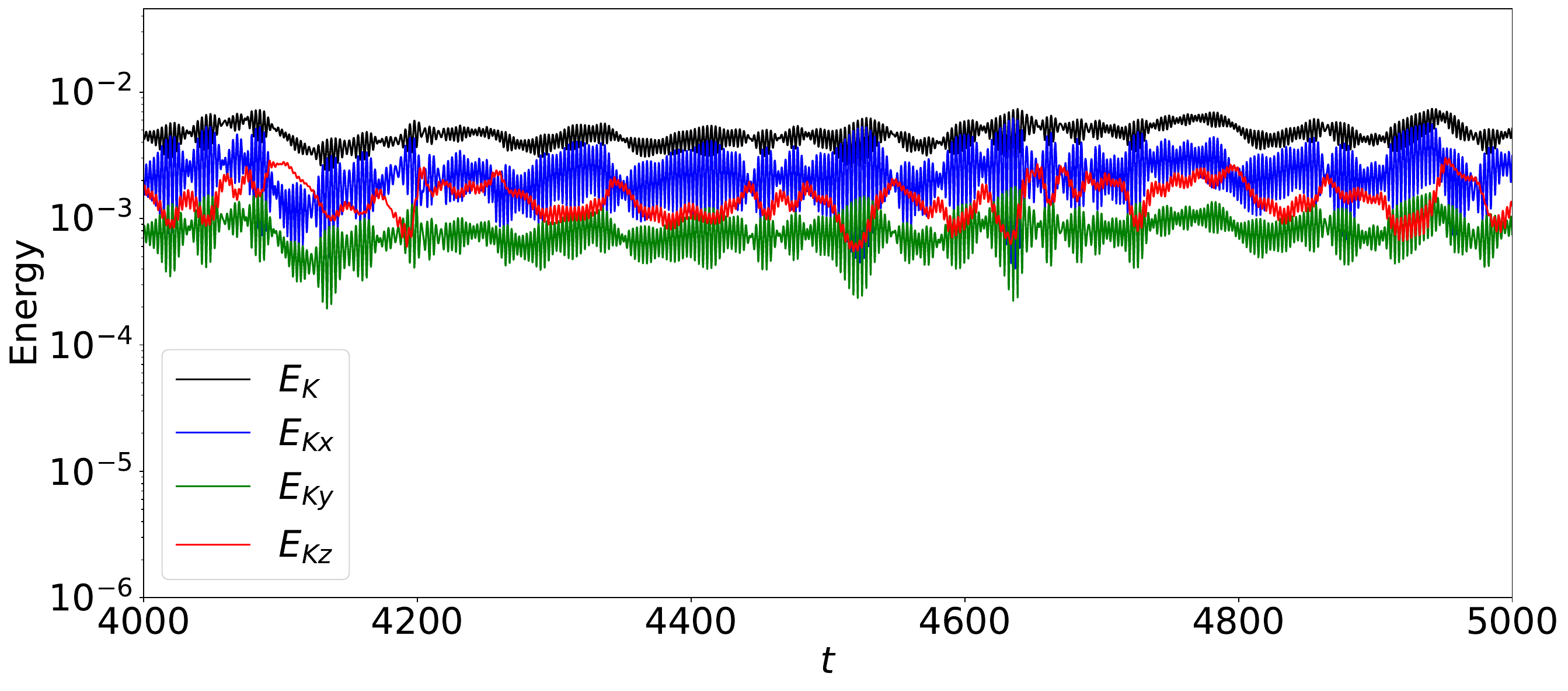}
    (b)\hskip -2mm \hspace{2mm} \vspace{3mm}{\label{}\includegraphics[width=0.9\linewidth]{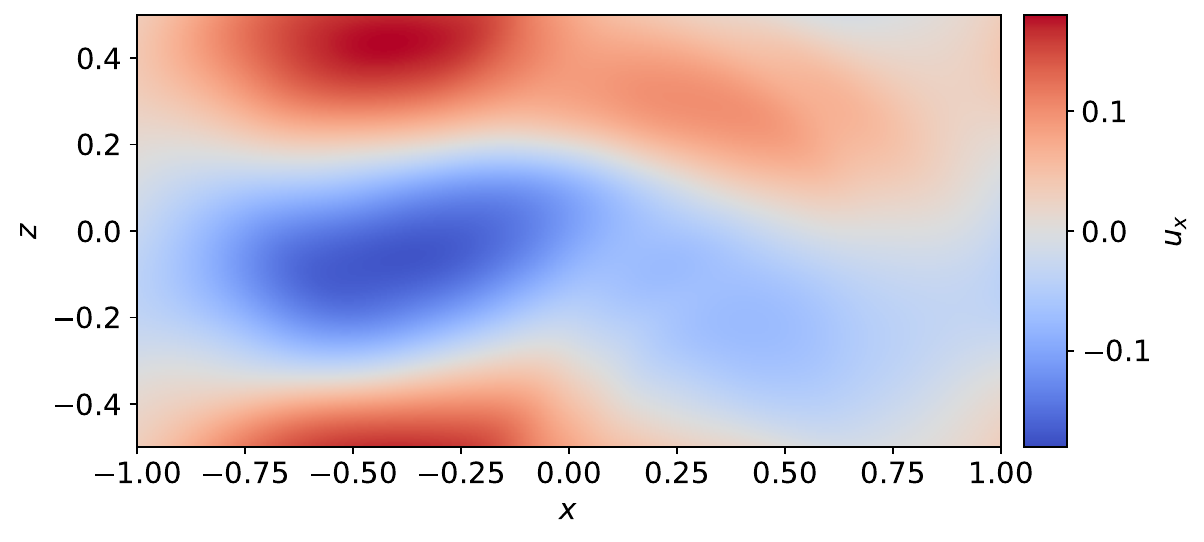}}
    (c)\hskip -2mm \hspace{2mm} \vspace{3mm}{\label{}\includegraphics[width=0.9\linewidth]{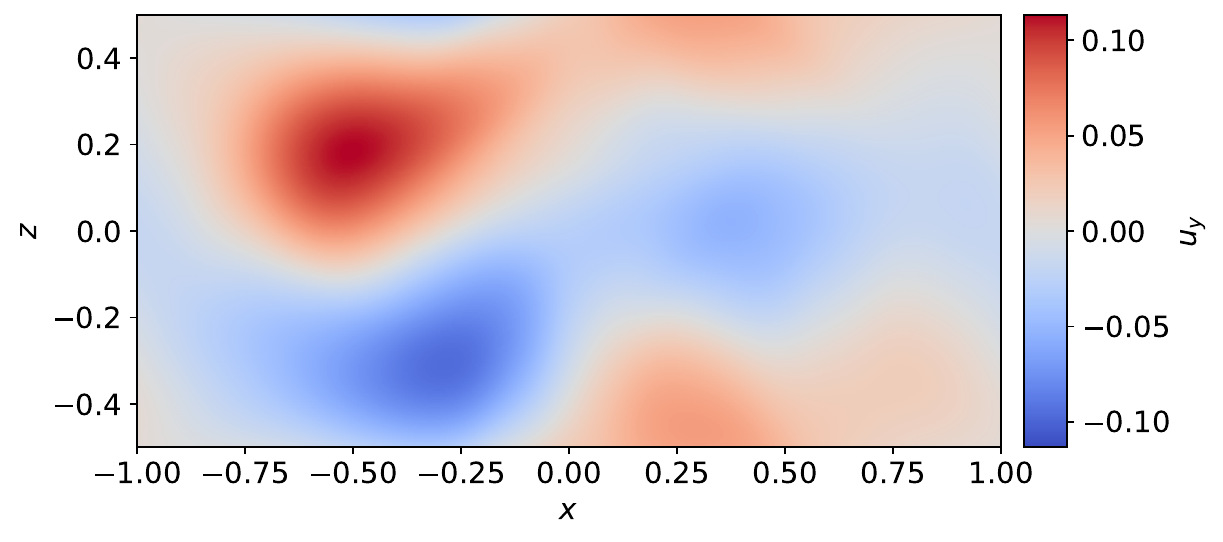}}
    (d)\hskip -2mm \hspace{2mm} \vspace{3mm}{\label{}\includegraphics[width=0.9\linewidth]{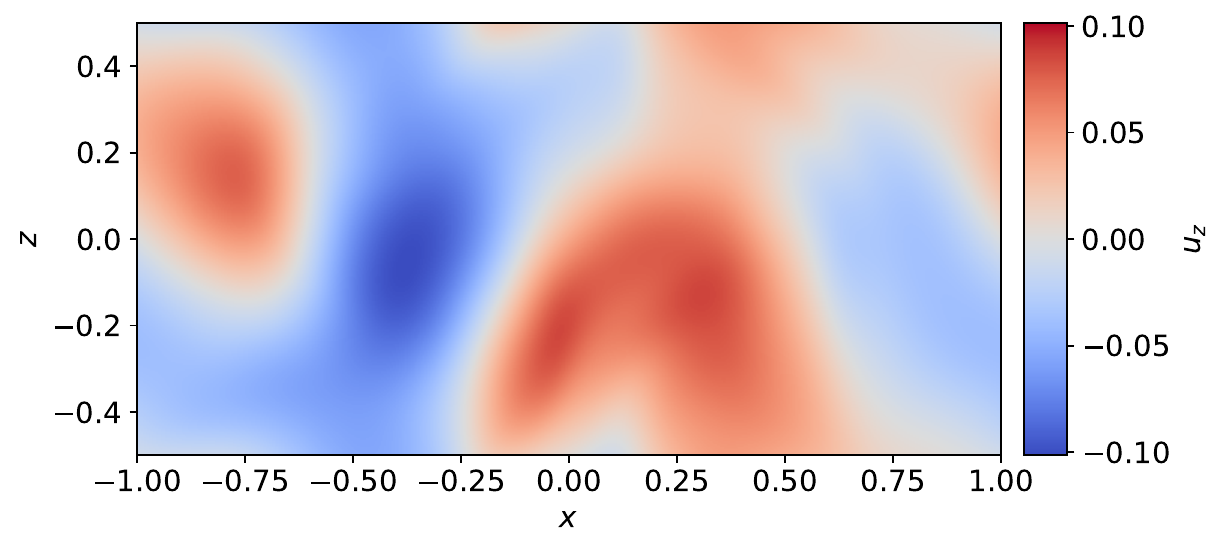}}

    (e)\hskip -2mm \hspace{2mm} {\label{}\includegraphics[width=0.42\linewidth]{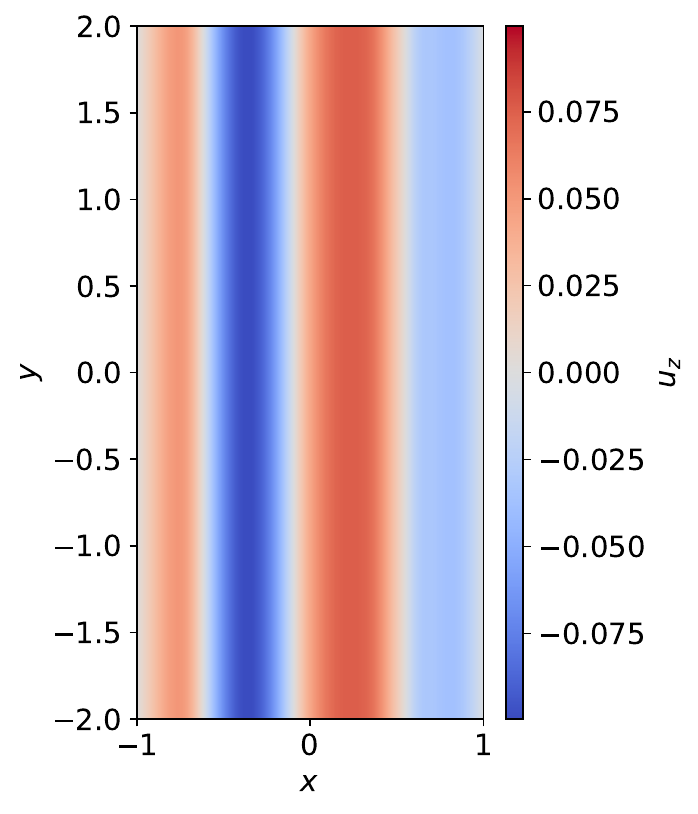}}
    (f)\hskip -2mm \hspace{2mm} {\label{}\includegraphics[width=0.4\linewidth]{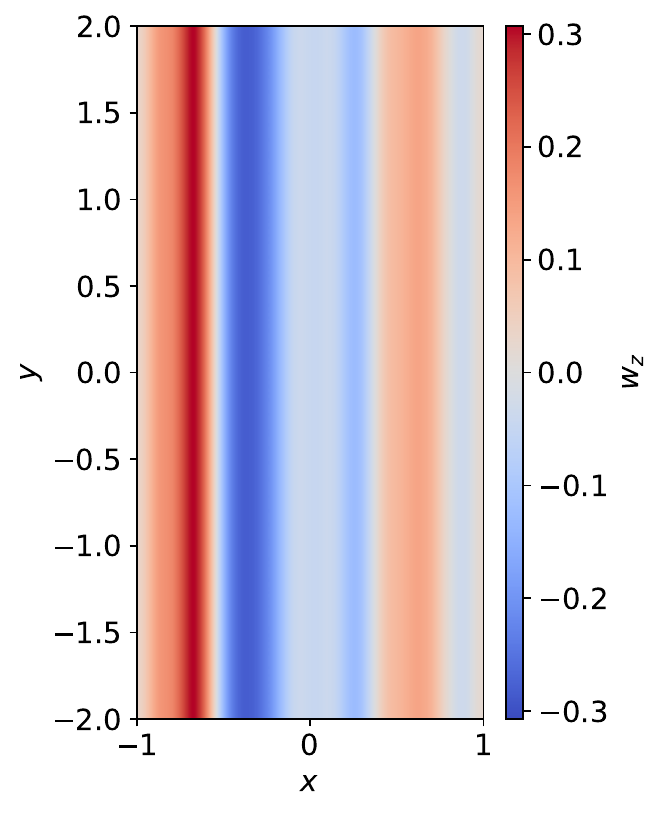}}
    \caption{Plots for a solution in the wave turbulent regime $(\text{Re}=10^{3.5}, \text{Pe}=4\pi^2)$. (a) Partial time series of the energy, and its directional parts. (b-f) Plots of components of the velocity field at $t=5.4\times10^3$: (b) $u_x$ in an $xz$-slice (at $y=0$); (c) $u_y$ in an $xz$-slice (at $y=0$); (d) $u_z$ in an $xz$-slice (at $y=0$); (e) $u_z$ in an $xy$-slice (at $z=0$); (f) $w_z$ in an $xy$-slice (at $z=0$).
    }
    \label{fig:WT}
\end{figure}

The dominance of a small number of modes seen in the WNL regime is replaced here with a more disordered state characterised by travelling inertial waves and a more chaotic evolution of the kinetic energies (Figs.~\ref{fig:WT}(a-d)). Nevertheless, this regime remains highly axisymmetric (Figs.~\ref{fig:WT}(e-f)) and is hence qualitatively equivalent to the regime found in axisymmetric simulations (TL21). 

The development of strong axisymmetry after a long period of pre-saturation growth raises the question of whether such states are stable to non-axisymmetric disturbances. In order to check this we perturbed the state shown in Fig.~\ref{fig:WT} with white noise at a magnitude similar to the existing axisymmetric velocity and continued the simulation. This did not trigger the emergence of non-axisymmetric structures and the strongly axisymmetric state re-emerged after approximately 40 time units.

\begin{figure}
    \centering
    (a)\hskip -2mm \hspace{2mm} \vspace{3mm}\includegraphics[width=0.9\linewidth]{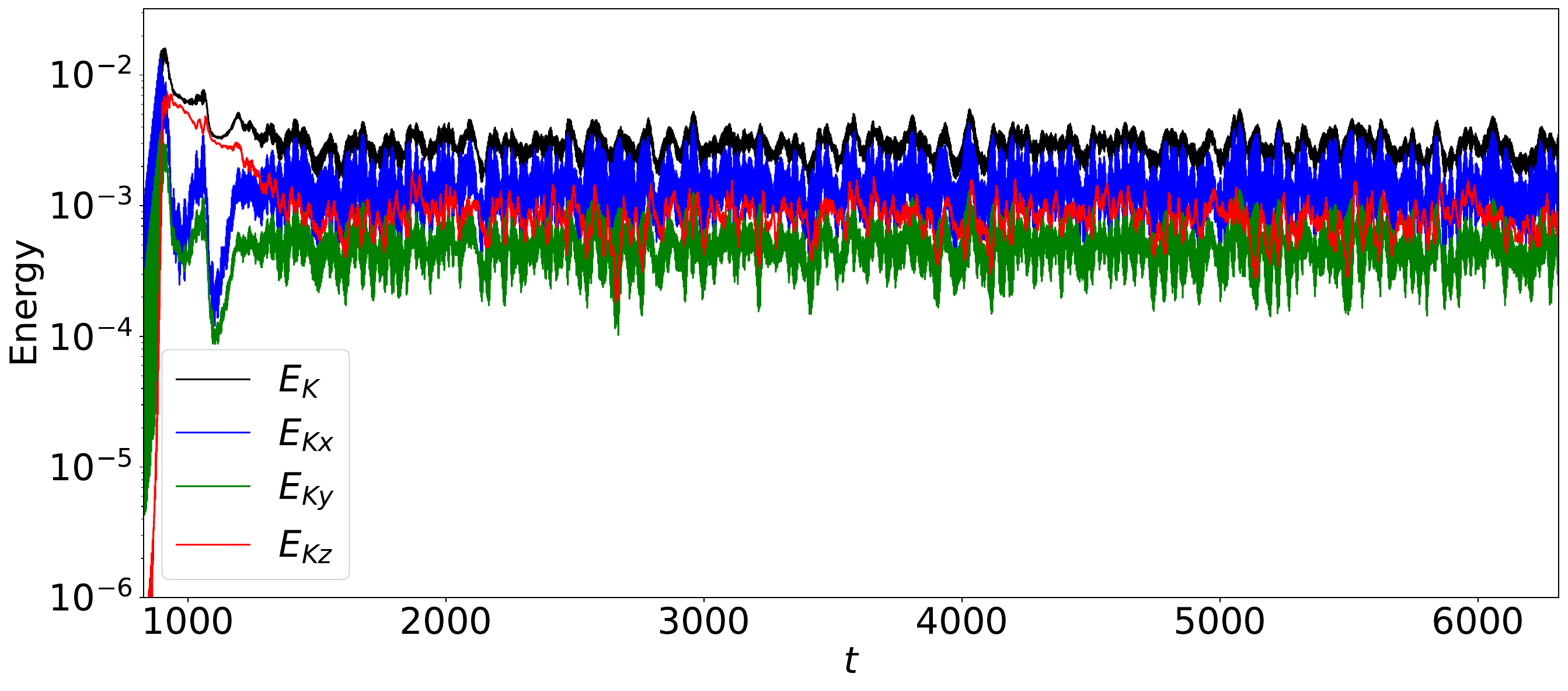}
    (b)\hskip -2mm \hspace{2mm} \vspace{3mm}{\label{}\includegraphics[width=0.9\linewidth]{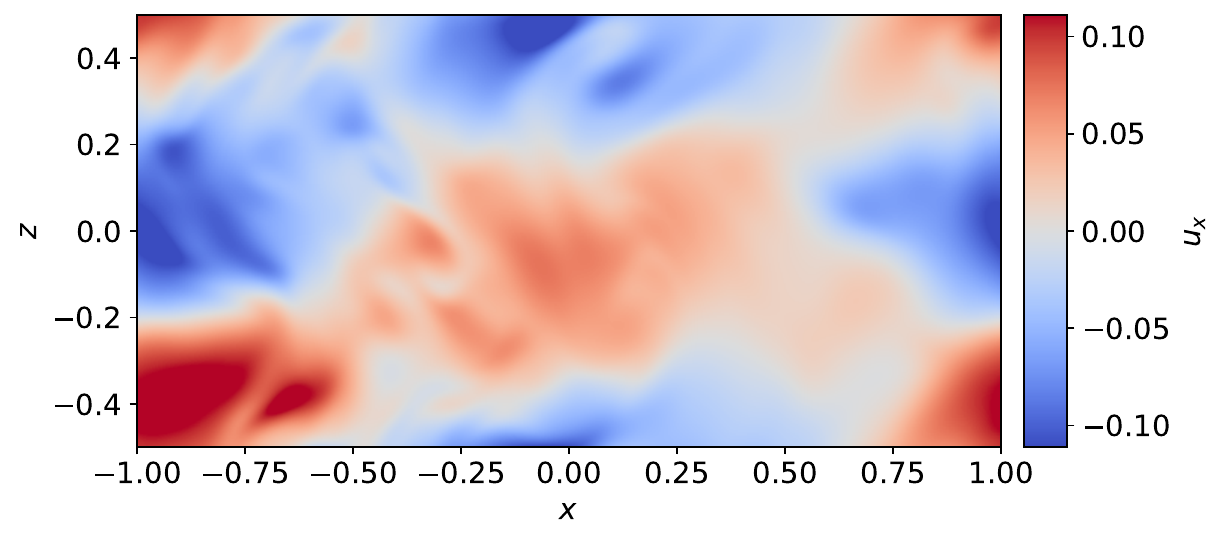}}
    (c)\hskip -2mm \hspace{2mm} \vspace{3mm}{\label{}\includegraphics[width=0.9\linewidth]{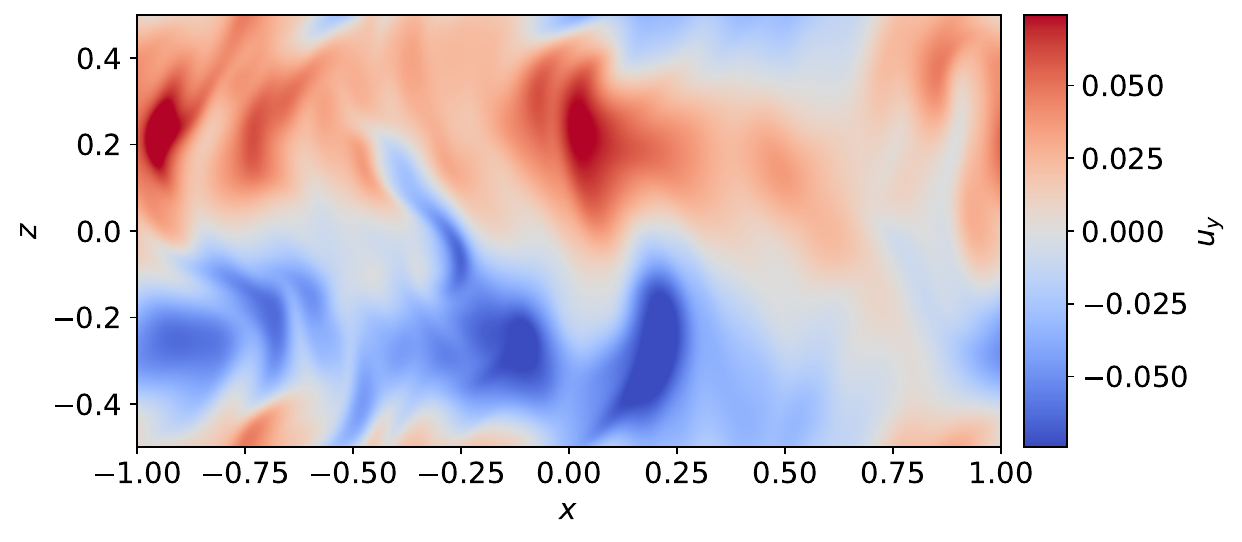}}
    (d)\hskip -2mm \hspace{2mm}\vspace{3mm} {\label{}\includegraphics[width=0.9\linewidth]{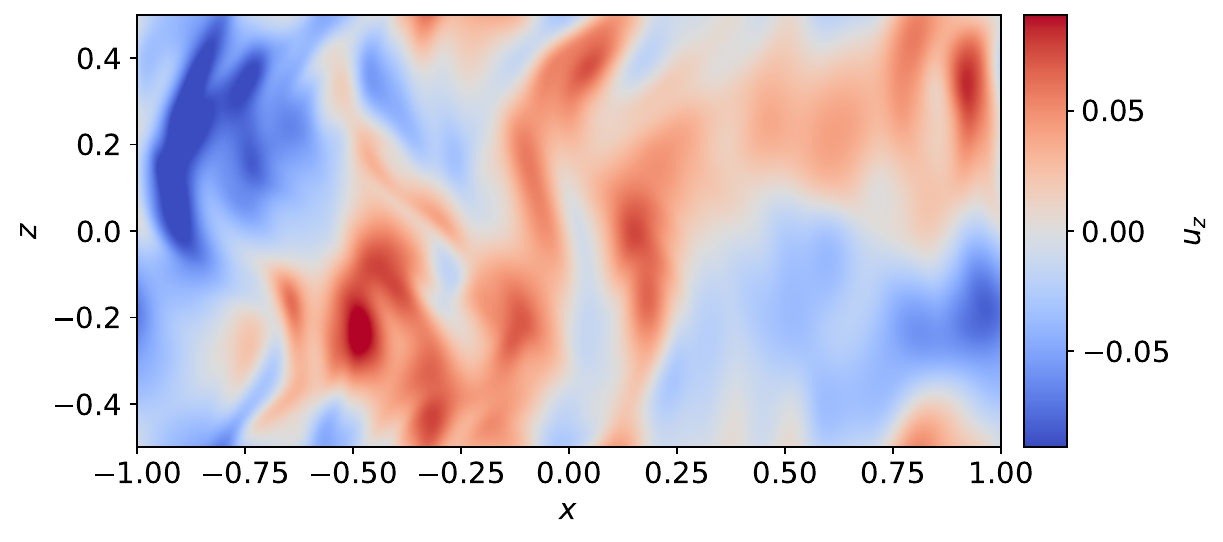}}

    (e)\hskip -2mm \hspace{2mm} {\label{}\includegraphics[width=0.42\linewidth]{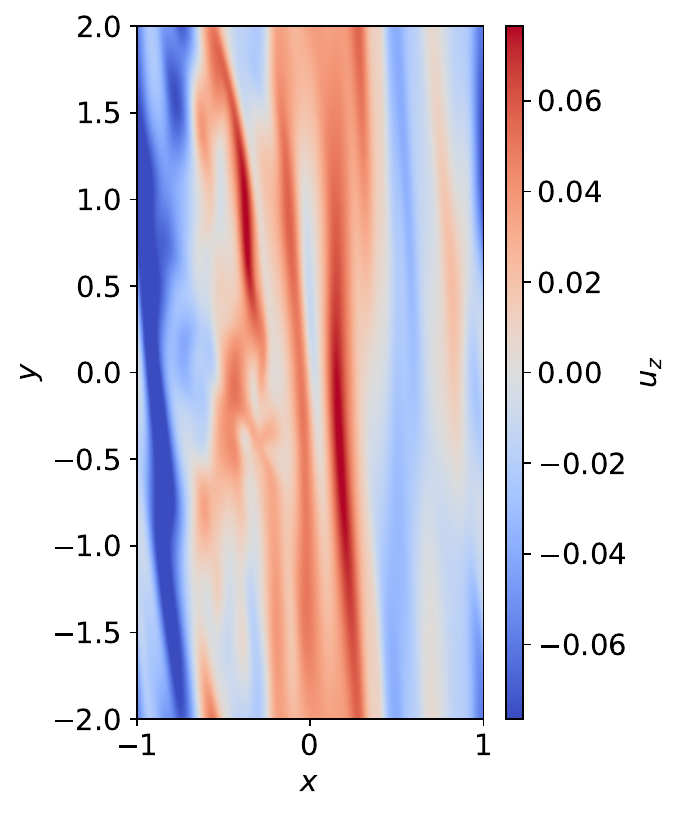}}
    (f)\hskip -2mm \hspace{2mm} {\label{}\includegraphics[width=0.4\linewidth]{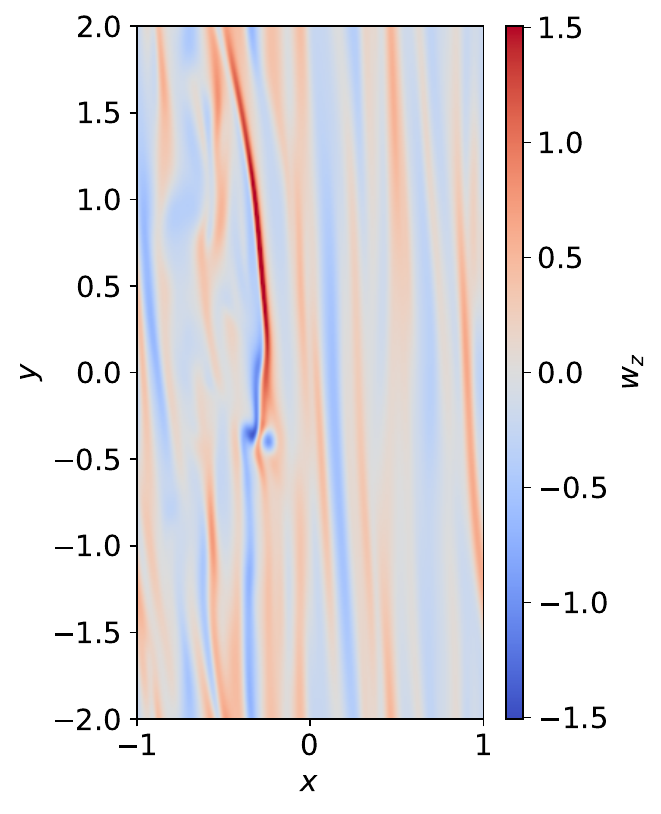}}
    \caption{Plots for a solution in the wave turbulent regime $(\text{Re}=10^{4}, \text{Pe}=4\pi^2)$. (a) Partial time series of the energy, and its directional parts. (b-f) Plots of components of the velocity field at $t=6.3\times10^3$: (b) $u_x$ in an $xz$-slice (at $y=0$); (c) $u_y$ in an $xz$-slice (at $y=0$); (d) $u_z$ in an $xz$-slice (at $y=0$); (e) $u_z$ in an $xy$-slice at ($z=0$); (f) $w_z$ in an $xy$-slice (at $z=0$).
    }
    \label{fig:WT2}
\end{figure}

At larger Reynolds number some intrinsic (unseeded) deviations from axisymmetry begin to appear (Fig.~\ref{fig:WT2}(e-f)), possibly due to nascent shear instability (see next subsection), and the flow becomes more turbulent (Fig.~\ref{fig:WT2}(b-d)). Lastly, the turbulent Reynolds stress is small and negative ($\sim 10^{-5}$), as in 2D (TL21).

\subsection{Cycles of zonal flows and vortices}

A further increase in the Reynolds number leads to
zonal flows emerging from the wave turbulence, completing the sequence of states found under axisymmetry (TL21).
Zonal flows are patterns in $u_y$ that form a banded structure in radius but are vertically and azimuthally homogeneous. Such flows are common in the geophysical context, where they form from a `geostrophic balance' between the Coriolis force and pressure gradient.
Within our simulation suite, we found zonal flows at Re$=10^{4.5}$ and Re$=10^5$ (for Pe$=4\pi^2$). Beyond the latter value of Re, due to the necessity of higher resolution, it becomes computationally prohibitive to investigate the parameter space in three-dimensions.

\begin{figure}
    \centering
    \includegraphics[width=0.9\linewidth]{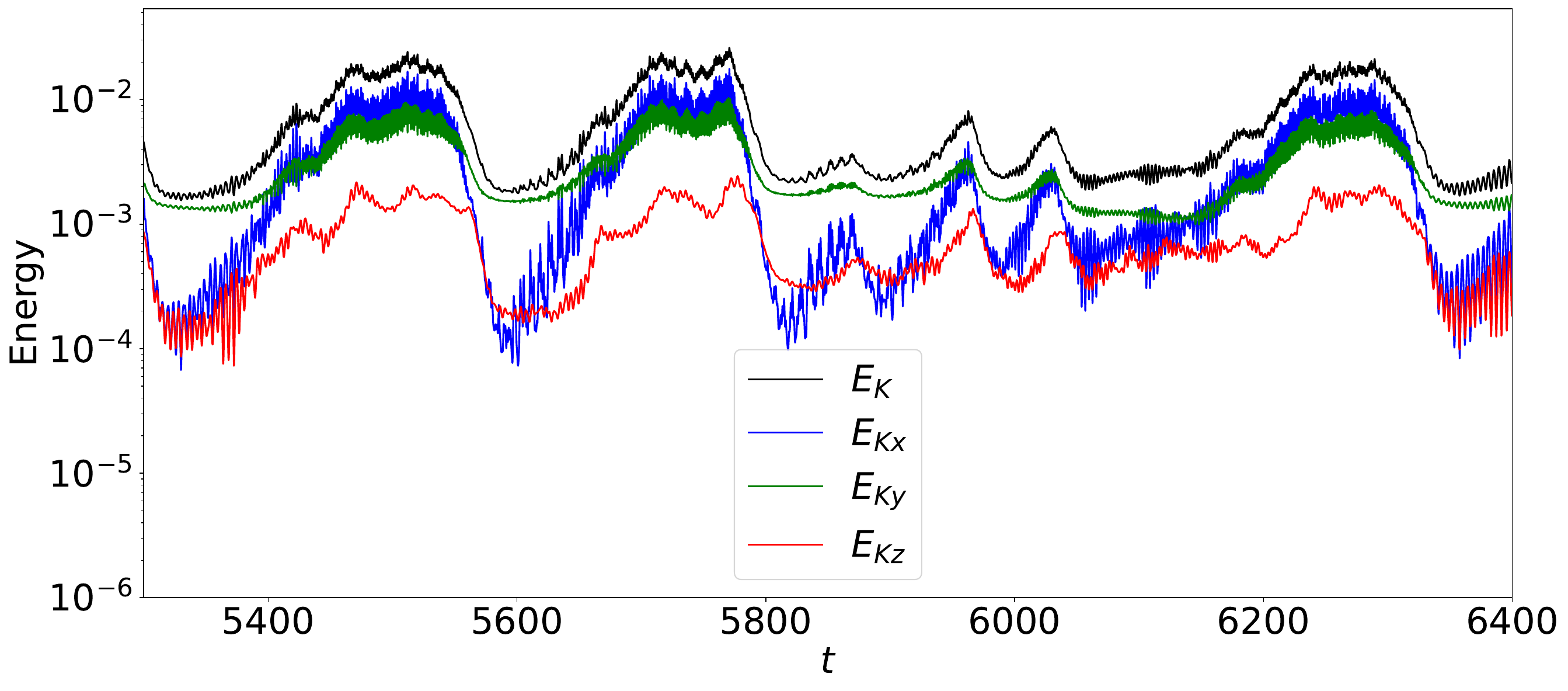}

    \caption{Partial time series of the energy, and its directional parts for a solution in the zonal flow/vortices regime $(\text{Re}=10^{4.5}, \text{Pe}=4\pi^2)$.
    }
    \label{fig:ZFV-energy}
\end{figure}

As was the case in axisymmetry, our 3D zonal flows first arise quasi-periodically in time, as evidenced in Fig.~\ref{fig:ZFV-energy} by the sequence of peaks and troughs in $E_{K_y}$. However, we find two key differences.

First, unlike the axisymmetric runs, which reverted to wave turbulence between zonal flows phases, in 3D runs vortices dominate these intervening periods. The azimuthal kinetic energy is highest during a phase with zonal flows (e.g.~$t\simeq6360$), whereas the radial and azimuthal energies are of a similar magnitude during phases with vortices (e.g.~$t\simeq6200$). Notably, the radial kinetic energy decays/grows by approximately two orders of magnitude during the transition to/from the zonal flow phase. Notably, despite the breakdown in axisymmetry, the mean Reynolds stress remains small and negative throughout the cycle.

Second, the strong elevator flows (banded structures in $u_z$, in the $xz$-plane) found in TL21's axisymmetric runs are no longer present. The vertical kinetic energy in our 3D runs is secondary throughout the time series, in stark contrast to TL21, where it provided the largest contribution to the total energy. It is possible that extending the domain in the azimuthal permits new parasitic modes varying in the $yz$-plane that stifle strong elevator flows. Nevertheless, during the vortex phase, the vertical energy does grow by approximately an order of magnitude (discussed later).

\begin{figure*}
    \centering
    \begin{minipage}{0.49\textwidth}   
    \centering
    (a)\hskip -2mm \hspace{2mm} \vspace{3mm}{\label{}\includegraphics[width=0.42\linewidth]{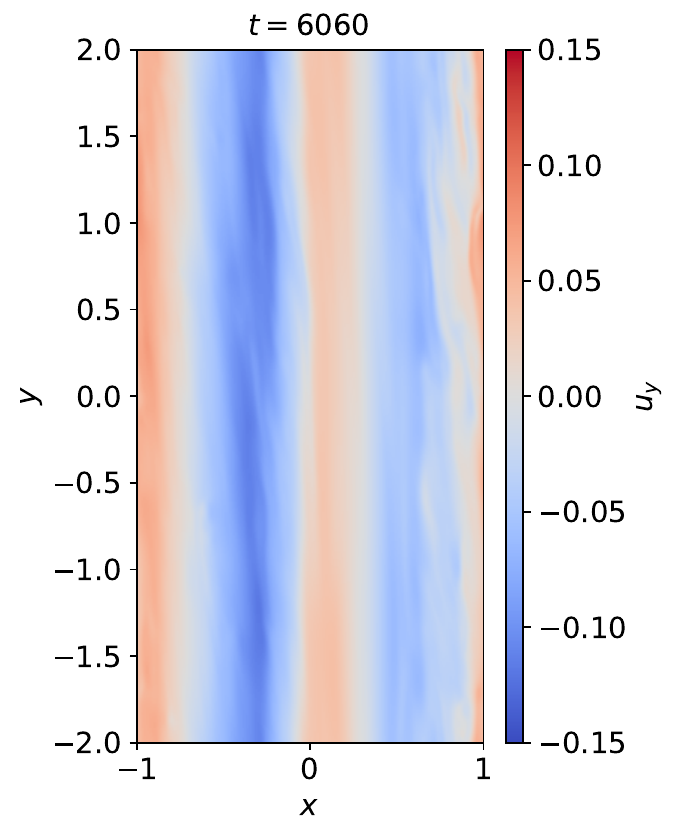}}
    (b)\hskip -2mm \hspace{2mm} {\label{}\includegraphics[width=0.42\linewidth]{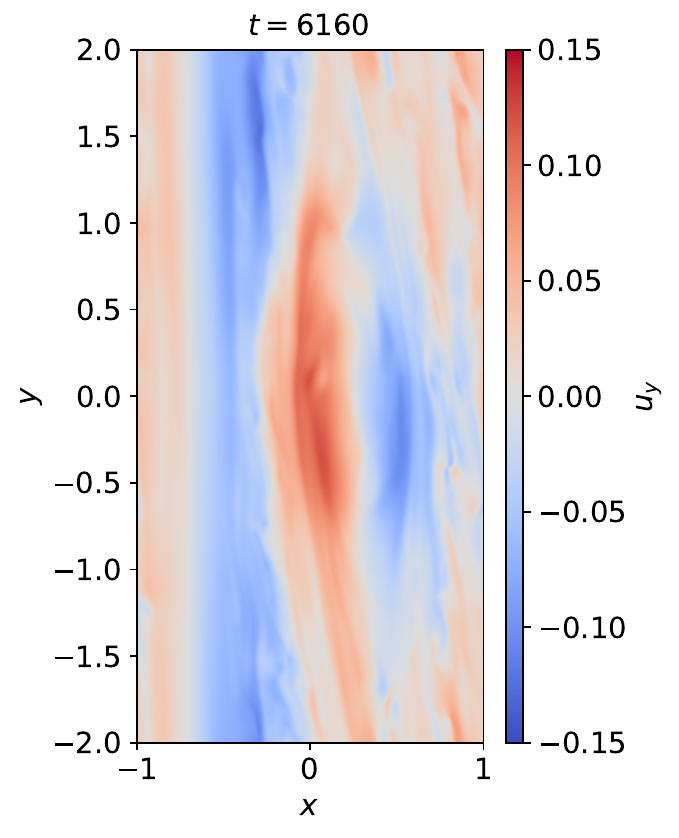}}
    
    (c)\hskip -2mm \hspace{2mm} \vspace{3mm}{\label{}\includegraphics[width=0.42\linewidth]{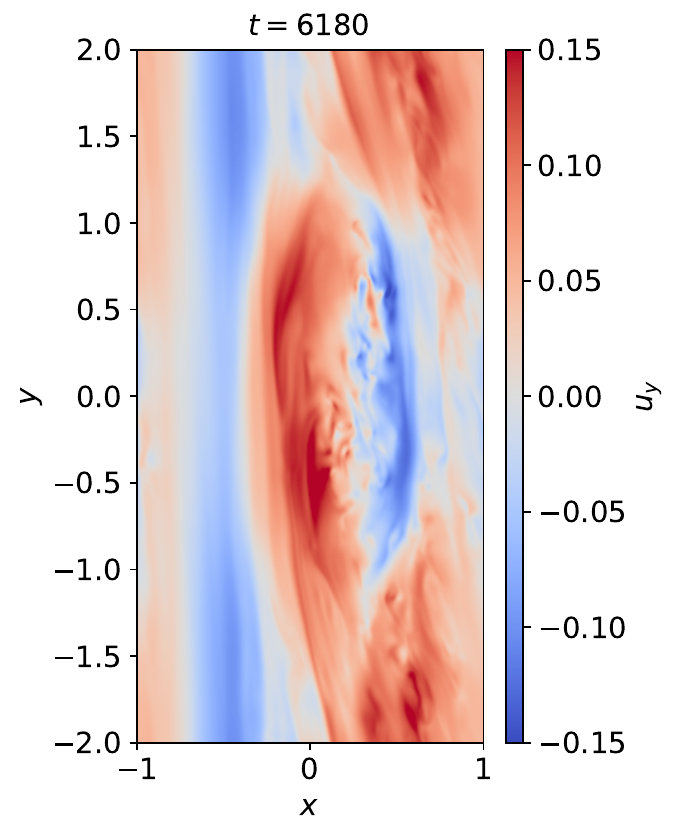}}
    (d)\hskip -2mm \hspace{2mm} {\label{}\includegraphics[width=0.42\linewidth]{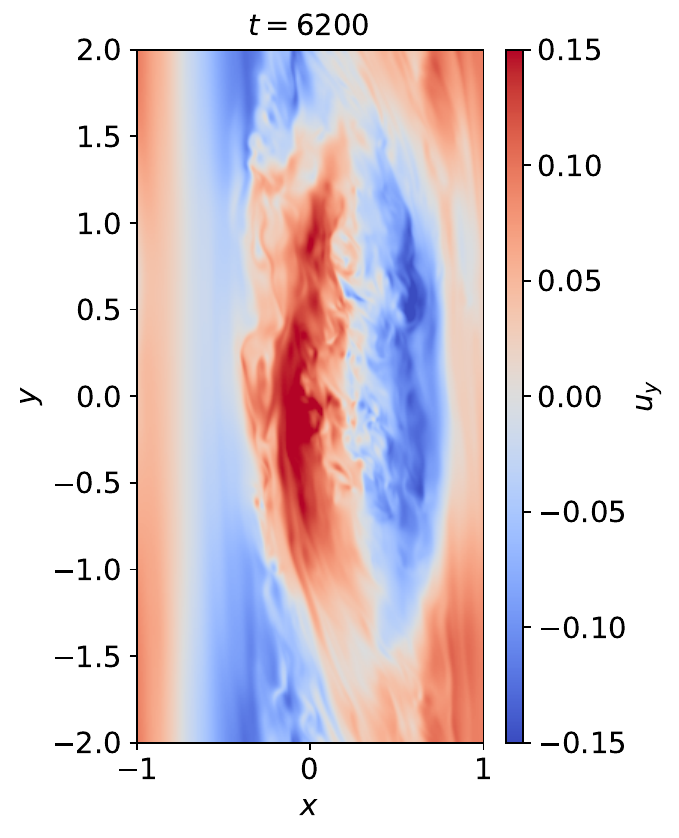}}
    
    (e)\hskip -2mm \hspace{2mm} {\label{}\includegraphics[width=0.42\linewidth]{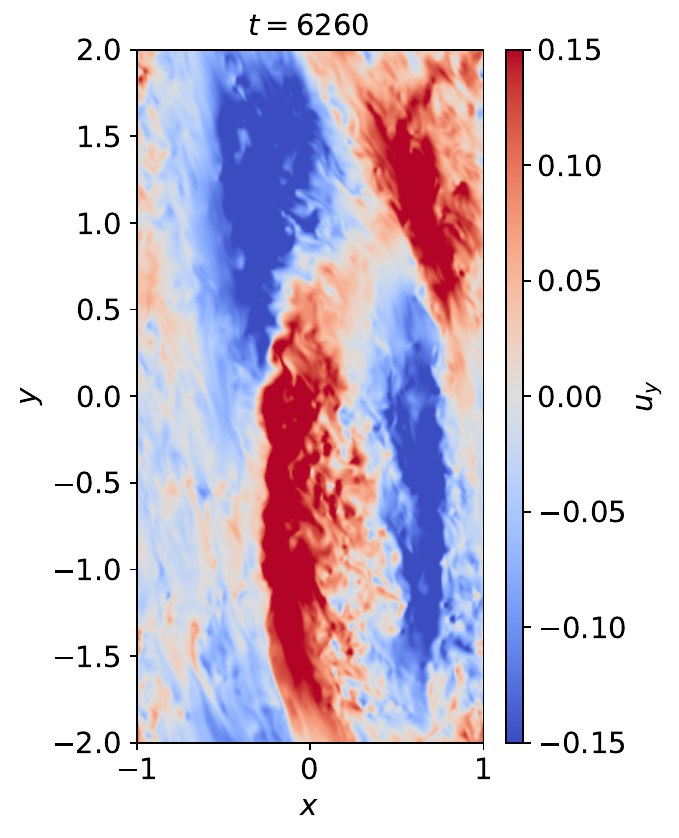}}
    (f)\hskip -2mm \hspace{2mm} {\label{}\includegraphics[width=0.42\linewidth]{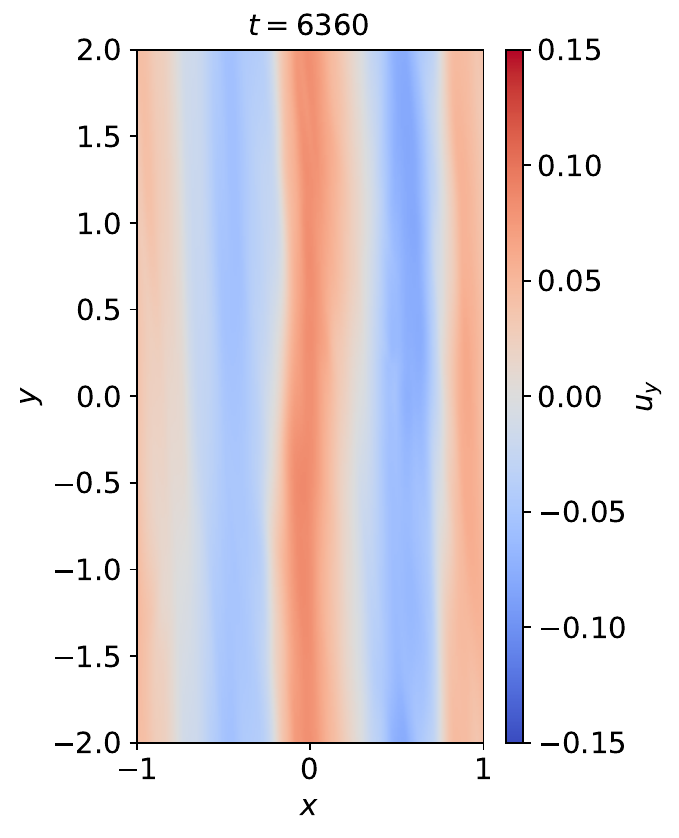}}  
    \end{minipage}
    \begin{minipage}{0.49\textwidth}
    \centering
    (g)\hskip -2mm \hspace{2mm} \vspace{3mm}{\label{}\includegraphics[width=0.41\linewidth]{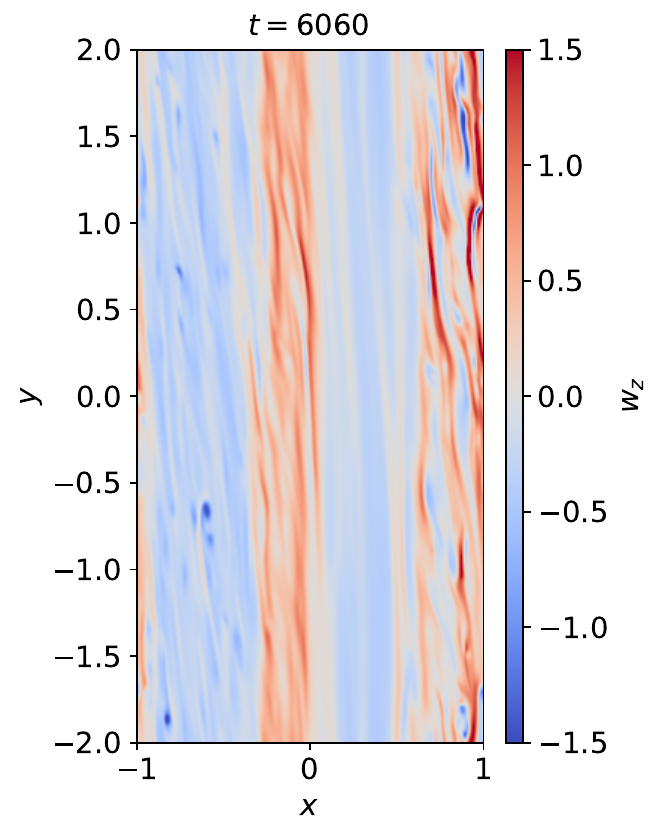}}
    (h)\hskip -2mm \hspace{2mm} {\label{}\includegraphics[width=0.41\linewidth]{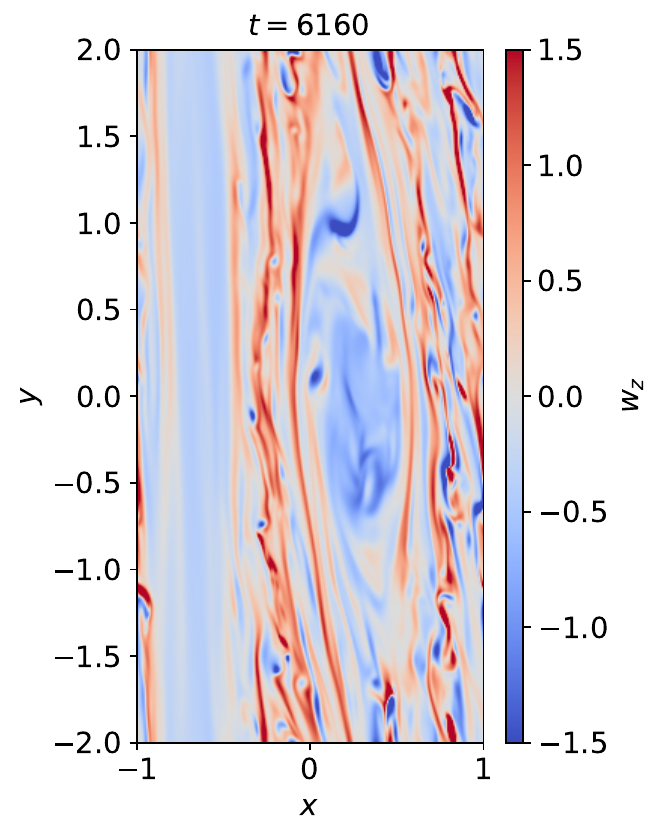}}
    
    (i)\hskip -2mm \hspace{2mm} \vspace{3mm}{\label{}\includegraphics[width=0.41\linewidth]{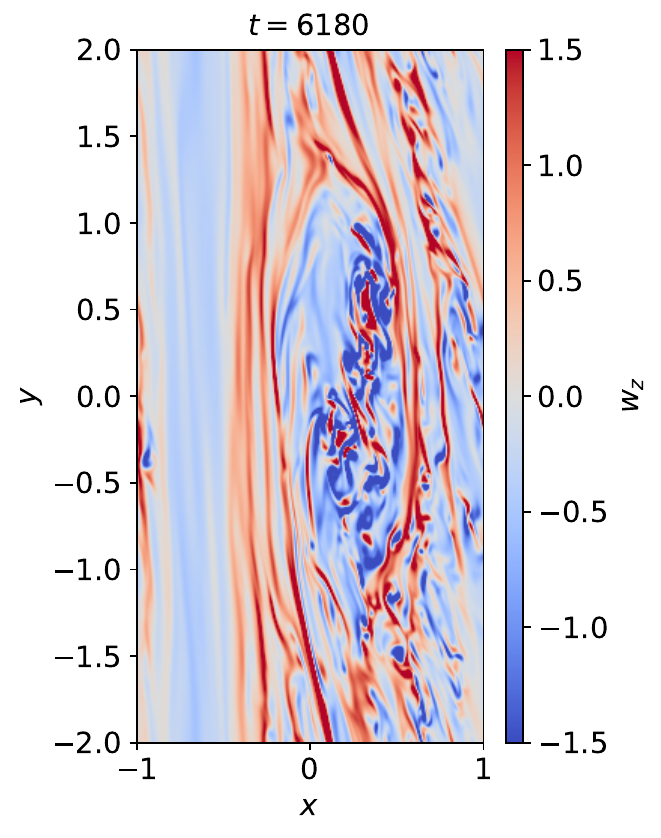}}
    (j)\hskip -2mm \hspace{2mm} {\label{}\includegraphics[width=0.41\linewidth]{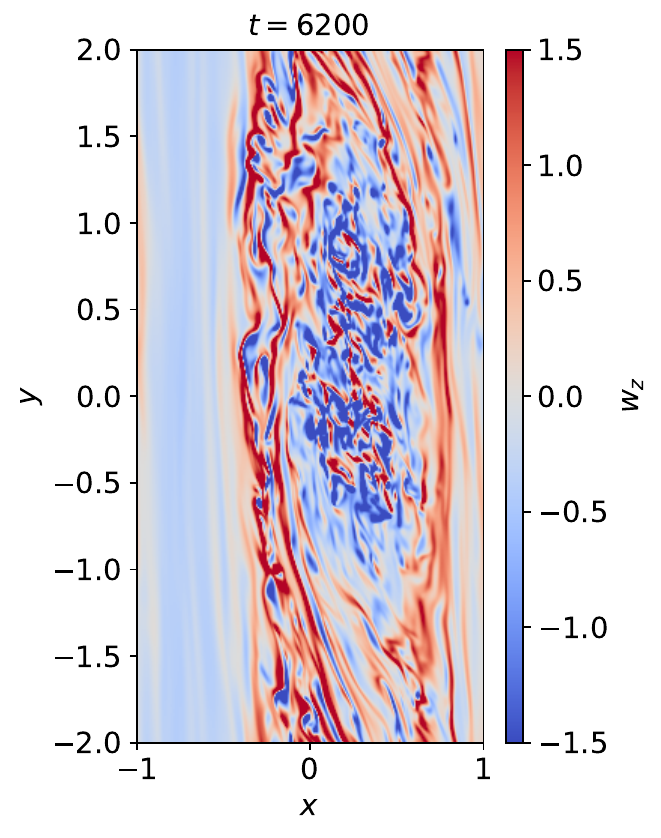}}
    
    (k)\hskip -2mm \hspace{2mm} {\label{}\includegraphics[width=0.41\linewidth]{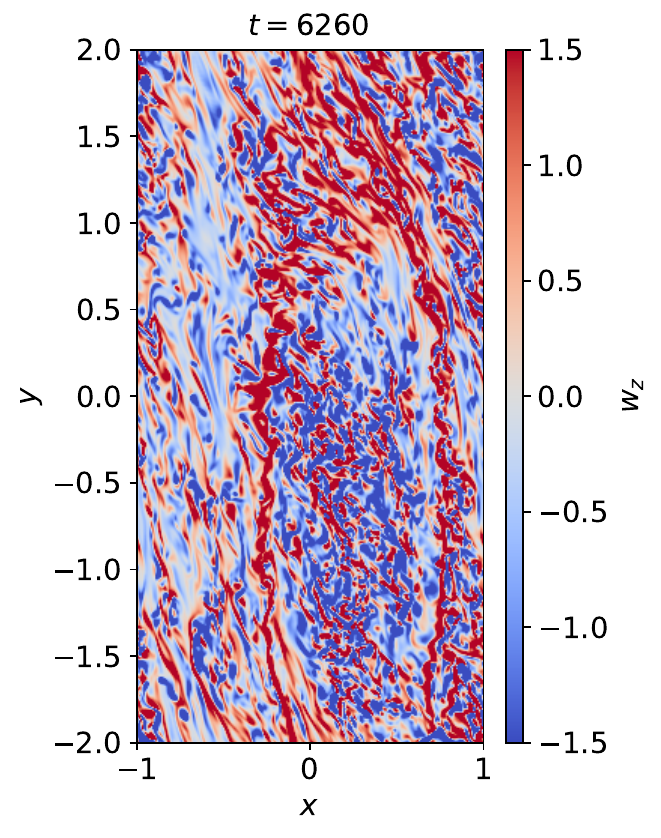}}
    (l)\hskip -2mm \hspace{2mm} {\label{}\includegraphics[width=0.41\linewidth]{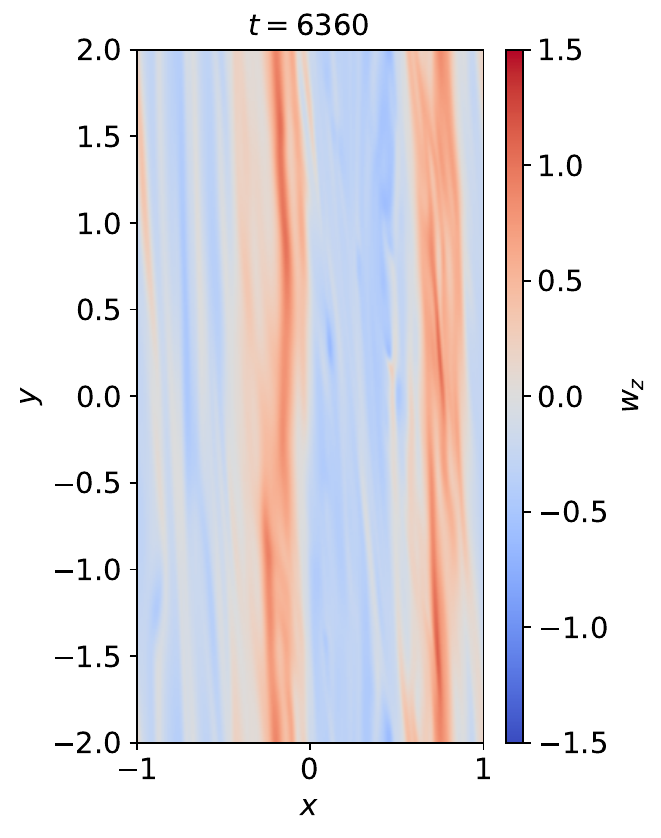}}
    \end{minipage}
    \caption{Time series of snapshots in the $xy$-plane (at $z=0$) of the azimuthal velocity and vertical vorticity for a solution in the zonal flow/vortices regime $(\text{Re}=10^{4.5}, \text{Pe}=4\pi^2)$. (a-f) Azimuthal velocity where (a) $t=6060$; (b) $t=6160$; (c) $t=6180$; (d) $t=6200$; (e) $t=6260$; (f) $t=6360$. (g-l) As (a-f) but for the vertical vorticity.}
    \label{fig:ZFV-series}
\end{figure*}

\begin{figure}
    \centering
    (a)\hskip -2mm \hspace{2mm} \vspace{3mm}{\label{}\includegraphics[width=0.9\linewidth]{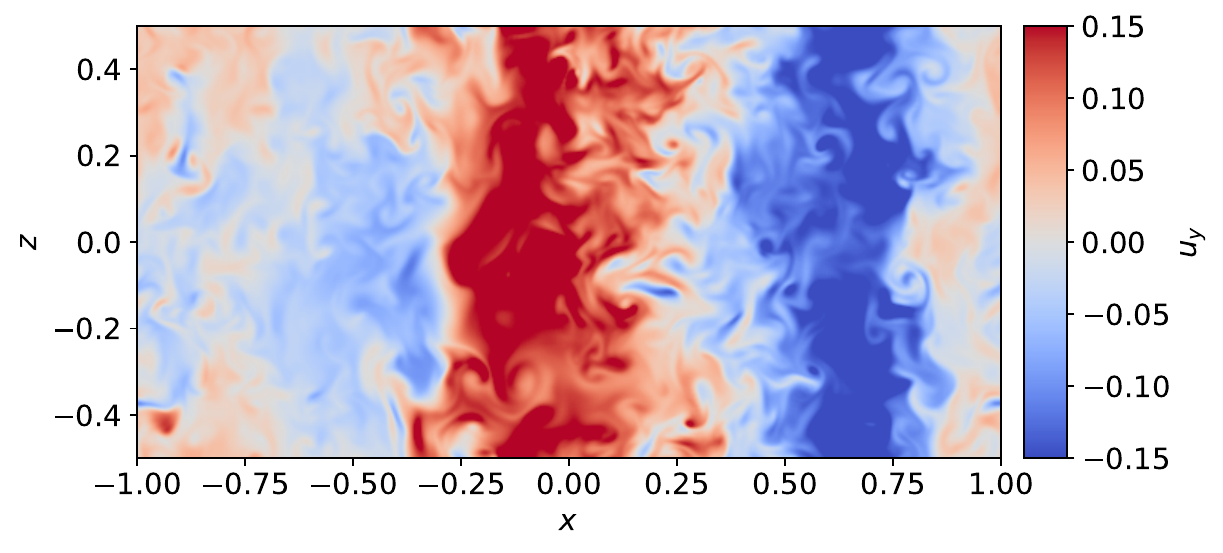}}
    (b)\hskip -2mm \hspace{2mm} \vspace{3mm}{\label{}\includegraphics[width=0.9\linewidth]{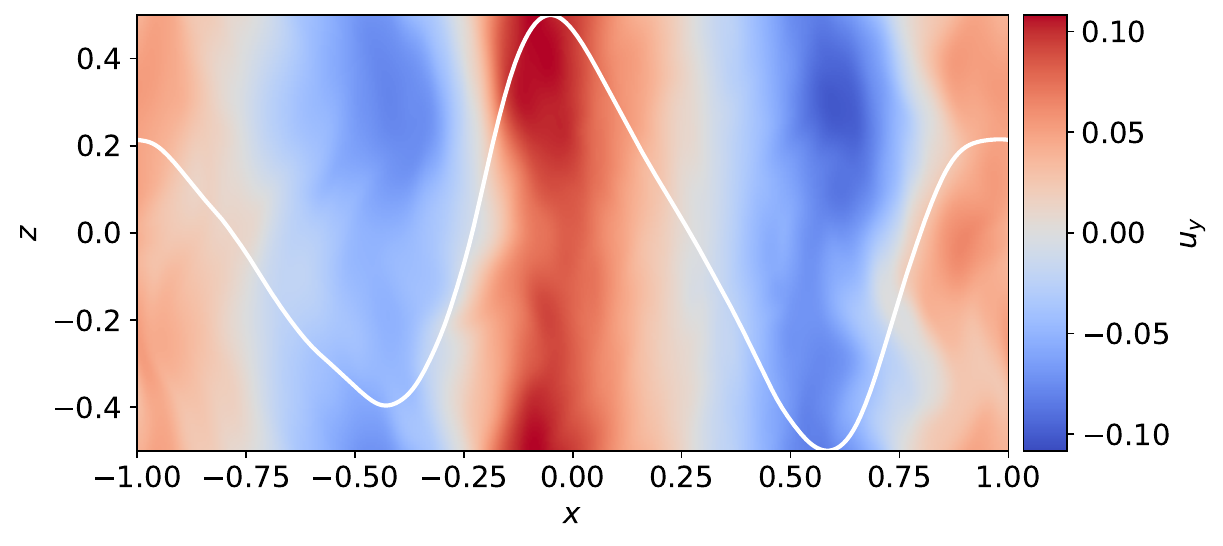}}
      
    (c)\hskip -2mm \hspace{2mm} {\label{}\includegraphics[width=0.9\linewidth]{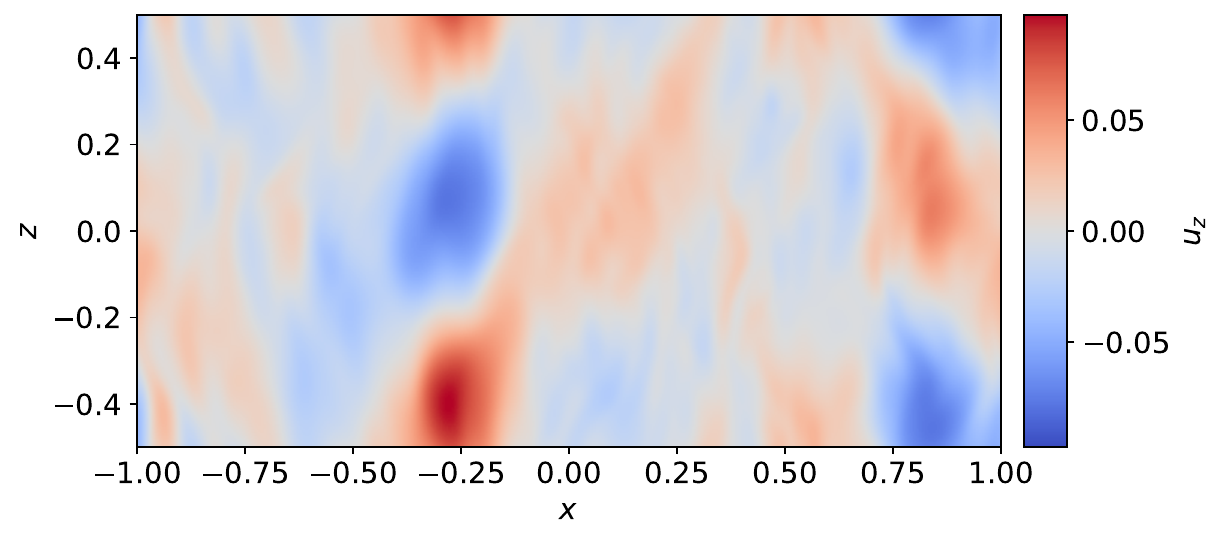}}

    \caption{Plots for a solution in the zonal flow/vortices regime $(\text{Re}=10^{4.5}, \text{Pe}=4\pi^2)$.
    (a) $u_y$ in an $xz$-slice (at $y=0$) at $t=6260$ (during the vortex phase); (b) $u_y$ in an $xz$-slice (at $y=0$) at $t=6360$ (during the zonal flow phase), overlaid with a white curve showing the profile of $u_y(x)$ averaged over both $y$ and $z$; (c) $u_z$ in an $xz$-slice (at $y=0$) at $t=6360$. 
    }
    \label{fig:ZFV}
\end{figure}

Fig.~\ref{fig:ZFV-series} shows the sequence of flow patterns during a typical cycle in the $xy$ plane. This sequence begins in a state dominated by strong zonal flows
and no clear vortex at $t=6060$: panels (a) and (g). Being periodic in radius, the zonal flow exhibits an inflexion point (observe the averaged $u_y$ profile in the middle panel of Fig.~\ref{fig:ZFV}) and is thus subject to shear instability in the $xy$ plane \citep[see, for e.g.,][]{Lith07,Van16,CY24}. Indeed, by $t=6160$ the zonal flow is disrupted, and a single anti-cyclonic vortex emerges of radial size $\sim 1$ and azimuthal size $\sim 3$: panels (b) and (h). Later, vortex sheets and small-scale vortical structures appear within it; panels (c) and (i). By $t=6260$, the zonal flows have also been fully suppressed in the $xy$-plane, whilst the large-scale vortex is violently destroyed. The relatively low aspect ratio ($\sim 3$) makes the vortex likely subject to the fast (centrifugal) branch of elliptical instability \citep{LP09, Rail14}, which possesses a growth rate $\lesssim \Omega$, consistent with our simulations. 
Finally, clear zonal flows re-emerge by $t=6360$, with no evidence of a strong vortex yet (panels (f) and (l)), and the cycle repeats.

Snapshots presented in Figs.~\ref{fig:ZFV} show flow patterns in the $xz$-plane from the same simulation. It is noteworthy that, although the axisymmetric nature of the zonal flows is destroyed during the vortex phase of the cycle (as seen in Fig.~\ref{fig:ZFV-series}), strong azimuthal flows do remain throughout, as observed in Figs.~\ref{fig:ZFV}(a-b). Hence the flow remains mostly independent of $z$ throughout the cycle, the main difference between the two phases being the small-scale structures seen on the fringes of the large-scale zonal flow in Fig.~\ref{fig:ZFV}(a). This is evidence of the elliptical instability at work as it breaks up vortices. In Fig.~\ref{fig:ZFV}(b), we also plot the $y$ and $z$ averaged $u_y$ (the overlaid white curve), which reveals a characteristic `sawtooth' pattern, with a characteristic shear layer width of $< 0.5$, notably less than the scale of the vortices appearing in Fig.~\ref{fig:ZFV-series}. 
Lastly, elevator flows, typified by $k=0$ structures in the $xz$-plane, are clearly absent (Fig.~\ref{fig:ZFV}(c)).

\begin{figure}
    \centering
     \includegraphics[width=0.9\linewidth]{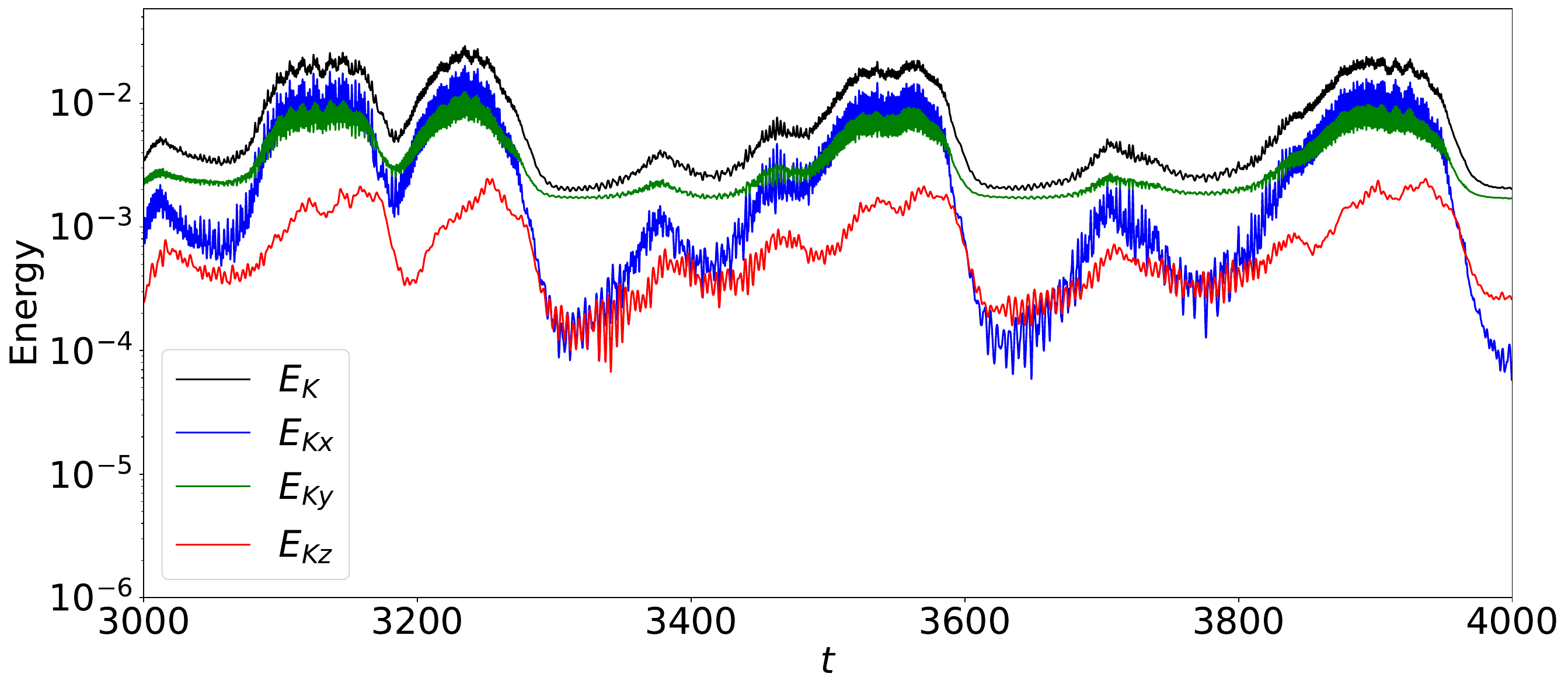}

    \caption{
    Partial time series of the energy, and its directional parts for a solution in the zonal flow/vortices regime $(\text{Re}=10^{5}, \text{Pe}=4\pi^2)$
    }
    \label{fig:ZFV2}
\end{figure}

The state containing cycles of zonal flows and vortices persists at the higher $\text{Re}=10^5$ (Fig.~\ref{fig:ZFV2}). The period of the cycle lengthens slightly compared with the case at lower Re (cf.~Fig.~\ref{fig:ZFV-energy}). During the zonal flows' destruction, the flow structures are more turbulent, though both phases of the cycle are very similar to those observed in Figs.~\ref{fig:ZFV-series} and \ref{fig:ZFV}. For this reason we do not display further plots from this simulation.

Finally, TL21's axisymmetric runs revealed that at large enough Re a state consisting of persistent zonal flows (i.e.~without cycles of wave turbulence or vortices) emerged. This state was not found in our current 3D simulation sequence at $\text{Pe}=4\pi^2$, which we were unable to extend beyond $\text{Re}=10^5$ due to computational restrictions. It is unclear, though perhaps unlikely, whether such a regime materialises for $\text{Re}>10^5$.

\subsection{The role of the SBI in zonal-flow/vortex cycles}

Until now, the SBI has not featured in our explanation of the sequence of COS states. The SBI is absent from the weakly nonlinear and wave turbulence states, which remain essentially axisymmetric and unable to offer vortical perturbations that the SBI could amplify. On the other hand, vortices did appear in the zonal-flow state, but they arise from straightforward shear instability acting on the zonal flows. While the SBI may not be responsible for the production of vortices, it might influence the evolution of vortices produced by shear instability, amplifying, reshaping, and resizing them. In fact, in so doing, the SBI may enhance a vortex's capacity to destroy the zonal flow that generated it.

\begin{figure}
    \centering
    (a)\hskip -2mm \hspace{2mm} {\label{}\includegraphics[width=0.42\linewidth]{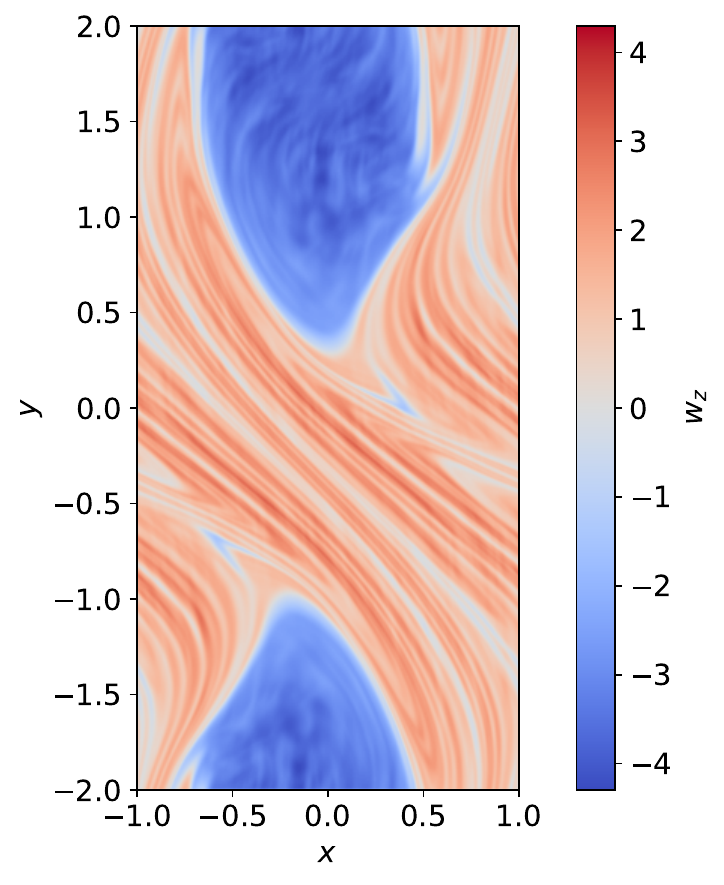}}
    (b)\hskip -2mm \hspace{2mm} {\label{}\includegraphics[width=0.42\linewidth]{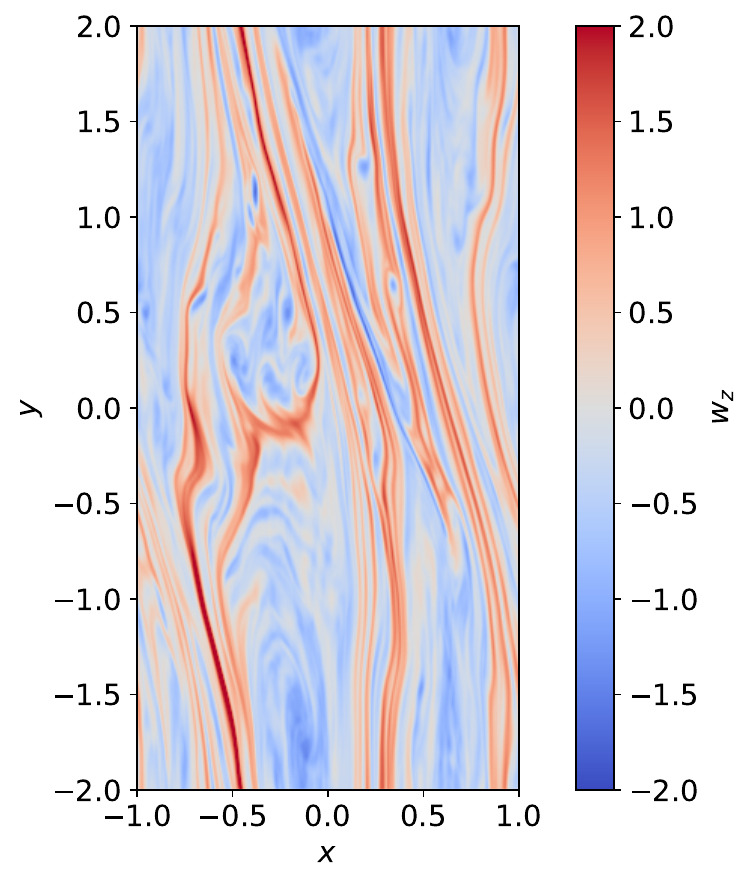}}
    
    \caption{Snapshots of the vertical vorticity for 2Dxy SBI simulations. (a) $\text{Re}=10^{4.5}, \text{Pe}=4\pi^2$. (b) $\text{Re}=4\times10^{5}, \text{Pe}=4000$.}
    \label{fig:SBI}
\end{figure}

First, to cleanly explore whether our 3D vortices could be influenced by the SBI, we performed 2D simulations under the same parameters, namely Re=$10^{4.5}$, Pe=$4\pi^2$. Fig.~\ref{fig:SBI}(a) demonstrates that a strong coherent vortex, elongated in the $y$-direction, emerges as a result of the SBI. The aspect ratio and azimuthal extent of the vortex are similar to those seen in Fig.~\ref{fig:ZFV-series}. Given that the vortical disturbance that seeded the vortex in both cases is different in 3D and 2D (non-axisymmetric shear instability versus large-scale noise, respectively), one might conclude that the SBI has resized and reshaped the initial 3D vortex. Indeed, between $t=6160$ and $t=6260$ (panels (h)-(j) in Fig.~\ref{fig:ZFV-series}), the vortex appears to grow in size, beyond the characteristic lengthscale of the linear instability. It is possible that the box size limits any further SBI-driven growth of the vortices in both 2D and 3D runs.

Second, we ran a 3D simulation with a large Pe for which we know the SBI is less effective (for given Re), so as to observe the behaviour of our COS vortices absent the SBI. As explained in LP10, the magnitude of thermal
diffusion (controlled via Pe) determines the size of the vortices that are preferred for amplification. At $\text{Pe}=4\pi^2$, the size of the vortices shed by COS-induced zonal flows appear sufficiently well matched to the SBI mechanism. We select now Pe=$4000$, for which the SBI prefers much smaller vortices.

\begin{figure}
    \centering
    (a)\hskip -2mm \hspace{2mm} \vspace{3mm}\includegraphics[width=0.9\linewidth]{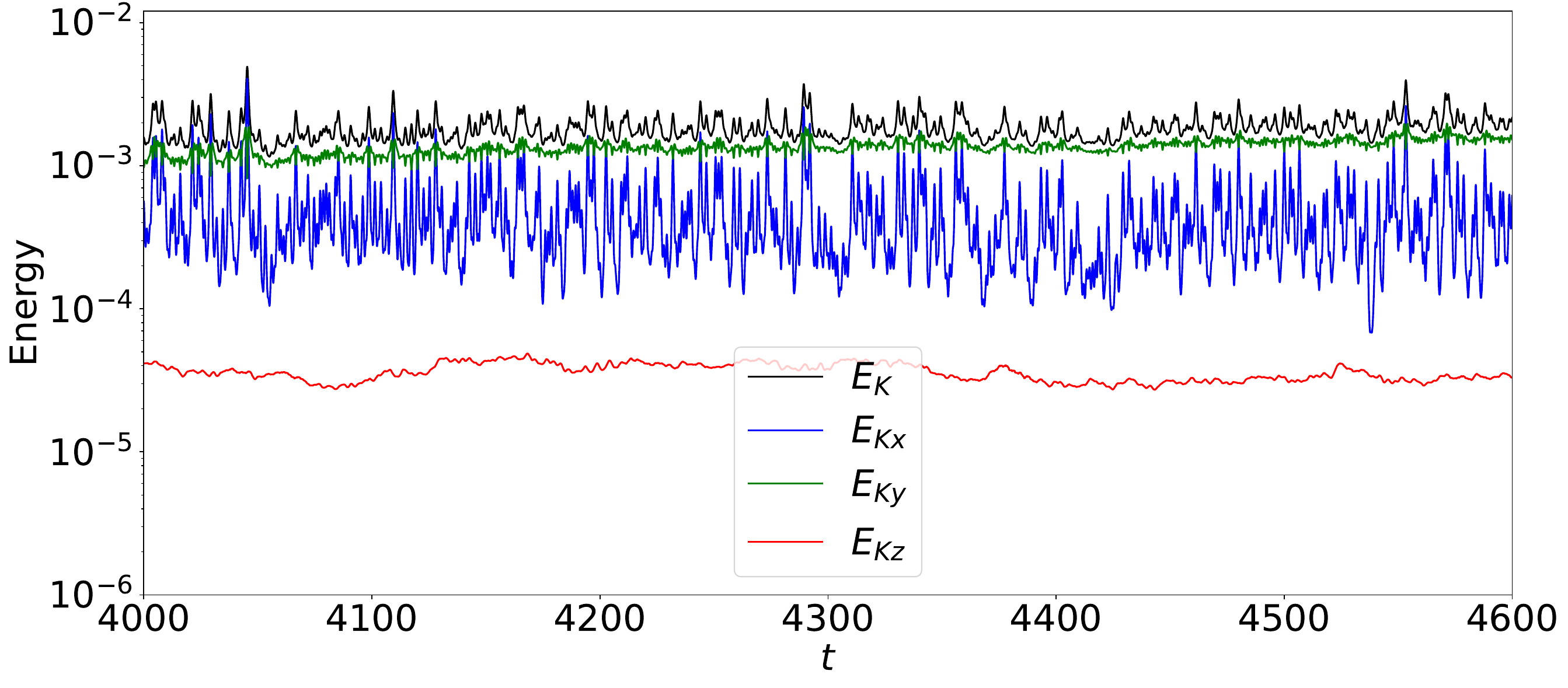}
    (b)\hskip -2mm \hspace{2mm} \vspace{3mm}\includegraphics[width=0.9\linewidth]{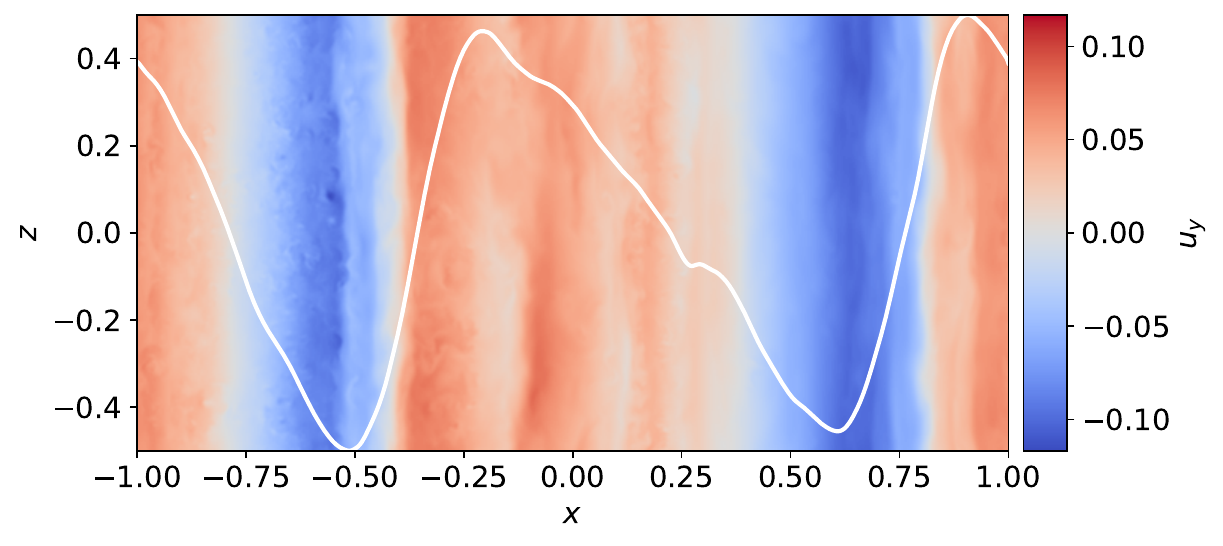}

    (c)\hskip -2mm \hspace{2mm} {\label{}\includegraphics[width=0.42\linewidth]{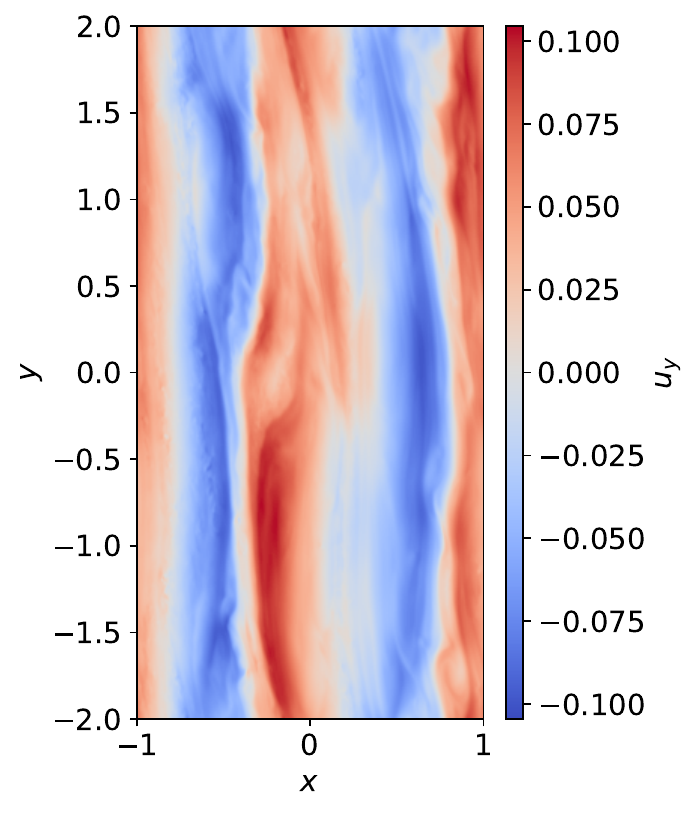}}
    (d)\hskip -2mm \hspace{2mm} {\label{}\includegraphics[width=0.4\linewidth]{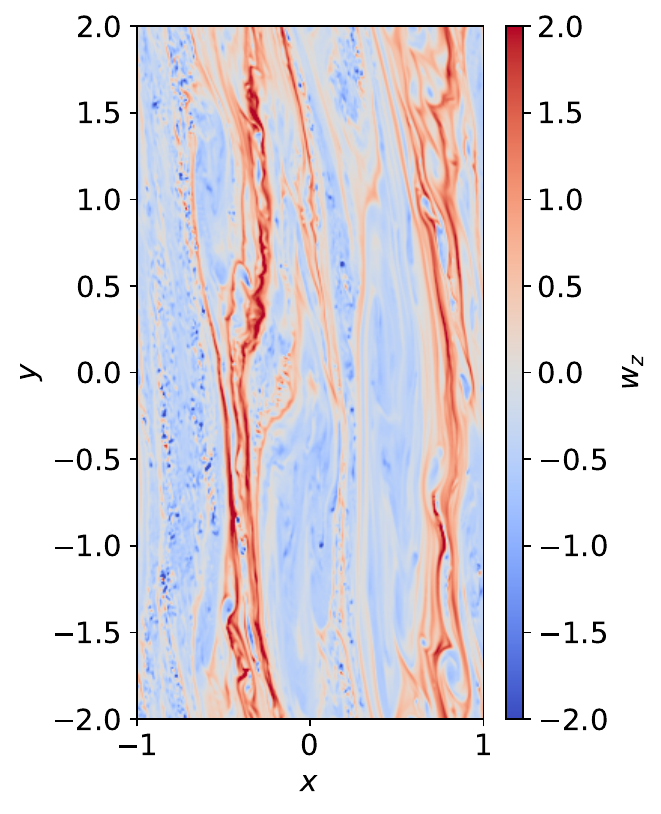}}

    \caption{Plots for a solution in the persistent zonal flows regime $(\text{Re}=4\times10^{5}, \text{Pe}=4000)$. (a) Partial time series of the energy, and its directional parts. (b-d) Plots of components of the velocity field at $t=4.5\times10^3$: (b) $u_y$ in an $xz$-slice (at $y=0$), overlaid with a white curve showing the profile of $u_y(x)$ averaged over both $y$ and $z$; (c) $u_y$ in an $xy$-slice (at $z=0$); (d) $w_z$ in an $xy$-slice (at $z=0$).
    }
    \label{fig:ZF}
\end{figure}

Fig.~\ref{fig:ZF} describes the state that arises for larger Pe. We find that zonal flows persist throughout the simulation. The zonal flow energy dominates at all times (Fig.~\ref{fig:ZF}(a)), in contrast to the intermittent zonal flow state where the radial energy came to dominate during periods of vortices (cf.~Fig.~\ref{fig:ZFV}(a)). The zonal flows may be seen in the snapshots in Figs.~\ref{fig:ZF}(b-c), and persist throughout the saturated state. The flows are strongly invariant in $z$; there is more variance in the azimuthal direction but the flows nevertheless form a clear banded structure, indicating they are zonal in nature. The flows are of a similar magnitude to their counterparts found during the cycles at lower Pe (cf.~Fig.~\ref{fig:ZFV}). The vorticity (Fig.~\ref{fig:ZF}(d)) is dominated by the zonal flows themselves, but does reveal additional sheets and weak vortices, in contrast to the regime at lower Pe where larger vortices emerged and grew in size and strength (cf.~Fig.~\ref{fig:ZFV-series}(h)).

To further characterise this regime, we conducted a 2D SBI run with the same large parameters. Fig.~\ref{fig:SBI}(b) shows a snapshot of the vertical vorticity in this run. As expected, the structures that develop are weaker, smaller, and far less coherent than the SBI simulation at lower Pe (cf. Fig.~\ref{fig:SBI}(a)). Clearly, the weakness of the SBI-amplification mechanism in this case means that, in 3D, vortices fail to fully disrupt the zonal flows and thus they persist.

\section{Discussion and conclusion}

We have presented a selection of numerical results that extend our previous 2D axisymmetric survey of COS states (TL21) to three-dimensions. The weakly nonlinear and wave turbulence states of TL21 remain essentially axisymmetric (and thus unchanged) in 3D, owing to the fact their characteristic timescales are much longer than the shear time. At lower Pe, the zonal flow states of TL21 are modified in 3D to produce vortices in the $xy$ plane, formed via shear instability. These are strengthened and enlarged by the SBI, break up the zonal flows, but are in turn destroyed by elliptical instability. Subsequently, the cycle of zonal-flow/vortex creation and destruction continues. 
At higher Pe, the SBI preferentially magnifies smaller vortices and, as a result of this mismatch, the shear-induced vortices are weaker and fail to break up the zonal flows, which now persist indefinitely. 

The implications of these results for dust accumulation by the COS in protoplanetary discs are significant. Depending on the efficiency of thermal diffusion, dust-capturing vortices are either strong, large, and intermittent (lasting a few tens of orbits) or rather weak and small. In neither case do COS-induced vortices present an assured route to planetesimal formation.  

While we feel our simulations are representative of a good portion of the COS's state space, and the relationship between the nonlinear development of the COS and the SBI, we have not conducted a comprehensive survey of the parameters, leaving that to future work. We must also point out the limitations of our physical model: it is local, Boussinesq, and adopts the diffusive approximation for radiative cooling. The latter assumption, in particular, limits our analysis to the optically thick regime, achievable perhaps near the midplane in the inner dead zone. Global effects, both radially and vertically are also likely to adjust the local dynamics described here, not least because the fastest growing COS modes possess a vanishing local radial wavenumber \citep[see also, e.g.,][]{klahr23,Lehm24}.
Finally, being Boussinesq, our simulations fail to capture the density waves that vortices naturally shed and their associated outward angular momentum transport. However, it is unlikely that they will interfere with the underlying cyclic dynamics we reveal, unless the energy loss via acoustic radiation is sufficient to oppose SBI-induced circularisation of vortices.

Direct comparison of our numerical results with previous work is difficult because of the (a) the large parameter space; (b) use of different radiation models, especially linear thermal relaxation, and (c) inclusion of extraneous physics, such as dust feedback. A linear cooling law, in particular, changes the nature of the SBI amplification mechanism: the SBI will preferentially select vortices of a certain aspect ratio (i.e., with turnover times matching the cooling time) not a certain size (as in this paper). 

The first point of comparison must be the small number of 3D simulations exhibited in LP10, which uses the same physical and numerical set-up and code. These numerical examples employ different parameters, Pe$=6400$ and $R=0.01$, and initial conditions, but nonetheless share some features with our runs, for instance cycles of vortex emergence and destruction, though a run begun from white noise produces a state of persistent vortices. The latter is indicative that the system might fall into one of several states, depending on the initial conditions. It is important to note, however, that for LP10's physical and numerical parameters the COS (and the dynamics it produces, zonal flows, etc.) is disfavoured; COS growth rates are 10 times smaller than in our simulations, and the fastest growing mode is only barely resolved.  

Local and global 3D compressible simulations of the COS have been undertaken by \citet{lyr14}, \citet{raet21}, \citet{lyra24}, and \citet{Lehm24, Lehm25}, with and without dust feedback, dust self-gravity, and the streaming instability. Importantly, these runs use a linear cooling law, not thermal diffusion. While the vortices that appear are seeded by the initial condition, they persist despite internal instabilites, and zonal flows do not emerge. One interpretation of these results is that, due to the selected cooling timescale, the SBI favours moderately elongated vortices, which are subject to only the weaker parametric branch of elliptical instability. The latter then permits the vortices to survive and to suppress any zonal flow development. 

To sum up our and others' work, in realistic discs at large enough Re, the COS supports a variety of dynamical states: (a) cycles of zonal flows and vortices, (b) persistent zonal flows (alongside weak vortices), and (c) persistent (albeit internally turbulent) vortices. This paper has highlighted the first two cases, while previous work has mainly revealed the third case. Further study is needed to sketch out the parameter boundaries between these outcomes, any dependence on the initial conditions, and (importantly) to better relate them to models with different treatments of radiative cooling. Once these dynamics are secure, we may then predict which COS states prevail in which PP disc regions, and consequently explore their role, if any, on planet formation.  

\section*{Acknowledgements}
The authors would like to thank the anonymous reviewer for a prompt report and a useful set of comments.

\section*{Data availability}

The data underlying this article will be shared on reasonable request to the corresponding author.

\bibliographystyle{mnras}
\bibliography{COS3D}

\end{document}